\documentclass[aps, prb, reprint, %
superscriptaddress, floatfix, longbibliography]{revtex4-2}

\usepackage{amsmath, amssymb, amsfonts, bm}
\usepackage{graphicx}
\usepackage{float}
\usepackage{color}
\usepackage[normalem]{ulem}
\usepackage[resetlabels]{multibib}

\usepackage[
colorlinks=true, linkcolor=blue, citecolor=blue, urlcolor=blue, 
setpagesize=false
]{hyperref}

\begin{document}

\title{Kekul\'{e} valence bond order in the Hubbard model on the honeycomb 
lattice with possible lattice distortions for graphene}

\author{Yuichi Otsuka}
\email{otsukay@riken.jp}
\affiliation{Computational Materials Science Research Team, 
RIKEN Center for Computational Science (R-CCS), 
Kobe, Hyogo 650-0047, Japan}
\affiliation{Quantum Computational Science Research Team, 
RIKEN Center for Quantum Computing (RQC), 
Wako, Saitama 351-0198, Japan}

\author{Seiji Yunoki}
\affiliation{Computational Materials Science Research Team,
RIKEN Center for Computational Science (R-CCS), 
Kobe, Hyogo 650-0047, Japan}
\affiliation{Quantum Computational Science Research Team, 
RIKEN Center for Quantum Computing (RQC), 
Wako, Saitama 351-0198, Japan}
\affiliation{Computational Condensed Matter Physics Laboratory, 
RIKEN, Wako, Saitama 351-0198, Japan}
\affiliation{Computational Quantum Matter Research Team, 
RIKEN Center for Emergent Matter Science (CEMS), 
Wako, Saitama 351-0198, Japan}

\date{\today}

\begin{abstract}
  
  We investigate if and how the valence-bond-solid (VBS) state emerges 
  in the Hubbard model on the honeycomb lattice when the Peierls-type
  electron-lattice coupling is introduced.
  We consider all possible lattice-distortion patterns allowed for this 
  lattice model for graphene which preserve the reflection symmetry 
  and determine the most stable configuration in the adiabatic limit 
  by using an unbiased quantum Monte Carlo method.
  The VBS phase with Kekul\'{e} dimerization is found to appear 
  as an intermediate phase between a semimetal and an antiferromagnetic 
  Mott insulator for a moderately rigid lattice.
  This implies that the undistorted semimetallic graphene can be driven 
  into the VBS phase by applying strain, accompanied by the single-particle excitation 
  gap opening.

\end{abstract}

\maketitle

\section{Introduction} 

Graphene is often described by the Hubbard model on a honeycomb lattice.
As a pioneering work on this model, Sorella and Tosatti elucidated 
a quantum phase transition from a semimetal (SM) to an antiferromagnetic Mott 
insulator (AFMI), which occurs at a finite strength of the interaction~\cite{Sorella-EPL1992}.
They further anticipated that their findings might be relevant for 
``the physics of strong correlations in the $\pi$-electron system in 
2D graphite''~\cite{Sorella-EPL1992}.
Later, after the synthesis of graphene~\cite{Novoselov-Science2004},
not only the peculiar noninteracting band structure~\cite{CastroNeto-RMP2009,DasSarma-RMP2011},
but also the many-body effects and their consequent quantum phase transitions 
in the Dirac electrons have been 
intensively studied~\cite{Kotov-RMP2012, Vafek-Review2014,Wehling-Review2014,Boyack-EPJ2021,Smith-IJMPA2021}.
Some of these studies were stimulated first by the possibility of a 
spin liquid phase~\cite{Meng-Nature2010,Chang-PRL2012,Seki-arXiv2012,Sorella-SciRep2012,Hassan-PRL2013,Chen-PRB2014,Ixert-PRB2014}
and later by an intriguing connection between the graphene physics 
and the celebrated Gross-Neveu model in high-energy 
physics~\cite{%
Herbut-PRL2006,Herbut-PRB2009a,Herbut-PRB2009b,Ryu-PRB2009,Assaad-PRX2013,ParisenToldin-PRB2015,Otsuka-PRX2016}.

While the effects of interaction on graphene have been to some extent 
understood at least based on lattice models, 
an experimental realization of AFMI in graphene, 
which would be highly promising for future device applications~\cite{Schwierz-NatNanotech2010},
has yet to be established. However, this does not necessarily prove that 
graphene is simply weakly correlated. Among many available estimations of the 
model parameters for graphene~\cite{Reich-PRB2002,Jung-PRB2013,Yazyev-PRL2008,Dutta-PRB2008,Wehling-PRL2011,Jung-PRB2011,Tang-PRL2015}, 
adopting the partially screened on-site Coulomb interaction of 
$U_{00} = 9.3 \mathrm{eV}$~\cite{Wehling-PRL2011}
and the widely accepted value of the hopping integral of 
$t \approx 2.7 \mathrm{eV}$~\cite{Reich-PRB2002,Jung-PRB2013}, 
we notice that their ratio is not far below the critical point of the Hubbard 
model on the honeycomb lattice 
$U_{\mathrm c}/t \simeq 3.8$%
~\cite{Sorella-SciRep2012,ParisenToldin-PRB2015,Otsuka-PRX2016,Seki-PRB2019,Raczkowski-PRB2020,Ostmeyer-PRB2021}.
This leads us to expect that AFMI is realized by applying strain,
since strain would reduce $t$ with less effect on the Hubbard interaction $U$.
It is noted that the application of strain has been explored as a way to introduce 
a gap in graphene not only by the Mott mechanism~\cite{Guinea-NatPhys2010,Choi-PRB2010,Cocco-PRB2010,Jiang-NanoLett2017,Si-Nanoiscale2016,Naumis-RepProgPhys2017}.

Recently, an \textit{ab initio} quantum Monte Carlo (QMC) calculation
examined the effect of strain on graphene and found that 
a Kekul\'{e}-like dimerized state, rather than AFMI,
is stabilized by increasing strain~\cite{Sorella-PRL2018}.
Being an insulator, this state should also be promising from an application point of view.
Theoretically, 
while the density functional theory calculations suggest the AFMI state 
as the ground-state of the strained graphene~\cite{Si-Nanoiscale2016,Lee-PRB2012},
it is fascinating that 
the more accurate description of the electron correlation, made possible by 
the QMC technique, predicts the Peierls-like ground state instead of the Mott insulator~\cite{Sorella-PRL2018}.

In this paper, as a complementary approach to the first-principles calculations,
we investigate the Kekul\'{e} valence bond order based on the simple 
lattice model, that is, the Hubbard model on the honeycomb lattice 
with electron-lattice (e-l) coupling, using the auxiliary-field quantum Monte Carlo
(AFQMC) method, which is numerically exact for the electron correlation.
We consider lattice displacements which modulate the electron hoppings 
and treat them adiabatically. 
Even with this simplification,
there may be a difficulty in determining the optimal bond-ordering pattern, 
as was the long-standing issue in similar models on a square 
lattice~\cite{Tang-PRB1988,Mazumdar-PRB1987,Ono-JPSJ2000,Chiba-JPSJ2003,Chiba-JPSJ2004a,Xing-PRL2021}. 
However, this difficulty is largely circumvented on the honeycomb lattice.
Frank and Lieb have shown that possible lattice distortions for graphene 
are periodic and reflection-symmetric, thus having at most six atoms 
per unit cell~\cite{Frank-PRL2011}.
Hence, together with the constant-volume condition, 
it is sufficient to consider two different lattice lengths in this unit cell.
We determine the optimal values of these lattice displacements by minimizing
the free energy at finite temperatures under a constant volume and thus obtain the phase diagram 
for several strengths of the e-l coupling,
which shows that the Kekul\'{e} valence-bond-solid (KVBS) phase is 
stabilized in the vicinity of $U_{\mathrm c}/t$.
For a parameter corresponding to a rigid lattice as in the case of graphene,
we find that the KVBS phase emerges as an intermediate phase 
between SM and AFMI, 
which suggests the correlation-driven SM-to-KVBS phase transition as predicted
by the preceding \textit{ab initio} QMC study~\cite{Sorella-PRL2018}.

The rest of the paper is organized as follows.
In the next section, we introduce the model and explain how to obtain the 
most stable bond configuration.
We then show the results of the QMC simulations, including the finite-temperature phase diagram, along with  
discussions in Sec.~\ref{sec:results_and_discussions}.
Finally, the summary is given in Sec.~\ref{sec:summary}.

\section{Model and method}
\label{sec:model_and_method}

 We study the e-l coupled Hubbard model described by the following 
Hamiltonian:
\begin{align}
 \mathcal{\hat{H}} =&
- \sum_{\langle i, j \rangle, \sigma} ( t - g u_{ij} ) \left(
\hat{c}_{i \sigma}^{\dagger} \hat{c}_{j \sigma} + \mathrm{h.c.}
\right) 
\nonumber \\
&+ U \sum_{i} \left( \hat{n}_{i \uparrow  } - \frac{1}{2}\right)
              \left( \hat{n}_{i \downarrow} - \frac{1}{2}\right)
+ \frac{K         }{2} \sum_{ \langle i, j \rangle } u_{i j}^{2},
\label{eq:hamiltonian}
\end{align}
where $\hat{c}_{i \sigma}$ annihilates an electron with spin 
$\sigma$ ($=\uparrow, \downarrow$) at site $i$, 
$t$ denotes the transfer integral in the absence of the
lattice distortion, 
$U$ is the Hubbard interaction, 
and $\hat{n}_{i\sigma}=\hat{c}^{\dagger}_{\sigma}\hat{c}_{\sigma}$ is a number operator.
The sum $\langle i, j \rangle$ runs over all nearest neighbor sites
of the honeycomb lattice.
The lattice displacement denoted by $u_{i j}$ is defined as 
a normalized relative change in the bond length;
the bond length between sites $i$ and $j$ is expressed as 
$a_{ i j } = a ( 1 + u_{ i j } )$,
where $a$ is the lattice constant of the undistorted lattice.
Thus, for example, 
if the bond shrinks, i,e, $u_{ij}<0$, the transfer integral increases 
by $ - g u_{ij}$ at the cost of the elastic energy
in the last term of Eq.~(\ref{eq:hamiltonian})
with $K$ being the elastic constant.

We consider the lattice degrees of freedom in the adiabatic limit,
assuming that the lattice displacements $u_{i j}$ are frozen at the
energy-minimizing configuration.
On the other hand, we take into account both quantum and thermal
fluctuations for the electron degrees of freedom.
Specifically, $u_{i j}$ is determined so as to satisfy the condition,
\begin{equation}
\frac{\partial F}{\partial u_{ij}} = 0, 
 \label{eq:adiabatic_approx}
\end{equation}
where $F=-\beta^{-1}\ln Z$ is the free energy at an inverse temperature 
$\beta=1/T$, and $Z$ is the grand-canonical partition function, i.e., 
$Z = \textrm{Tr}\ e^{-\beta \left( \mathcal{\hat{H}} - \mu \hat{N} \right)}$,
with $\hat{N}$ being the total number operator of electrons.
By setting the chemical potential $\mu=0$, we study the model at half filling.
The condition of Eq.~(\ref{eq:adiabatic_approx}) 
yields a self-consistent equation,
\begin{equation}
 \sum_{\sigma} g \langle c_{i \sigma}^{\dagger} c_{j \sigma} + \mathrm{h.c.} \rangle
+ K u_{i j} 
= 0.
\label{eq:self-consistent_equation}
\end{equation}
Here the bracket $\langle \cdot \rangle$ denotes calculating an expectation value,
for which we employ the finite-temperature version of AFQMC 
method~\cite{Blankenbecler-PRD1981,Hirsch-PRB1985,White-PRB1989,Assaad-CMP2008}.
Since the self-consistent equation is iteratively solved, 
the QMC simulation needs to be performed at each iterative step. 
Thus, in total, the required computational resources are rather large but feasible 
owing to the recent development of computational power. 
Several years ago, similar calculations were performed 
using the stochastic series-expansion QMC 
method~\cite{Sandvik-PRB1991,Sandvik-JPA1992,Sandvik-PRB1999}, 
which is less computationally demanding than AFQMC,
for the study of quasi-one dimensional molecular conductors~\cite{
Otsuka-JPSJ2008,Otsuka-Physica2009,Otsuka-Physica2012,Yoshioka-Crystal2012}.

\begin{figure}[tb]
  \centering
   \includegraphics[width=0.45\linewidth]{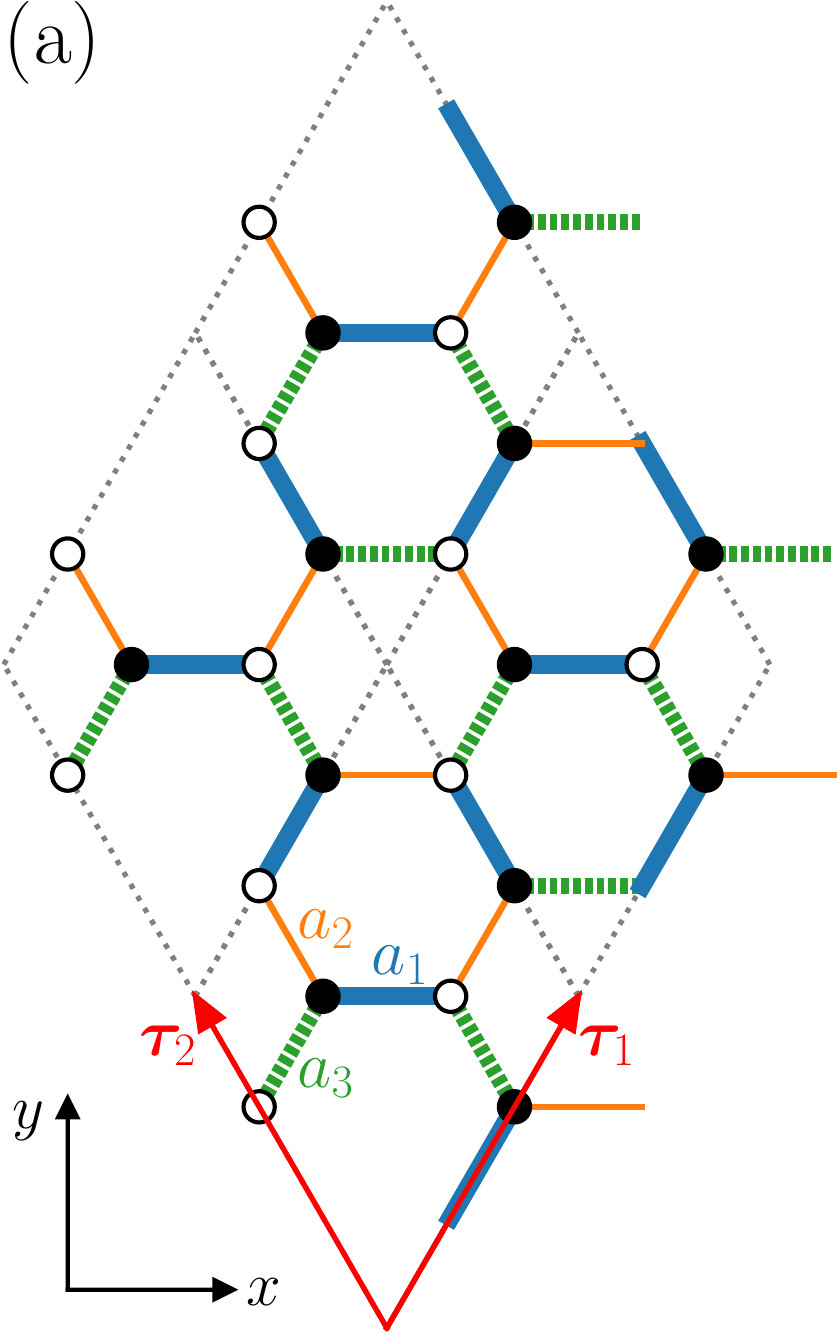}
  \includegraphics[width=0.45\linewidth]{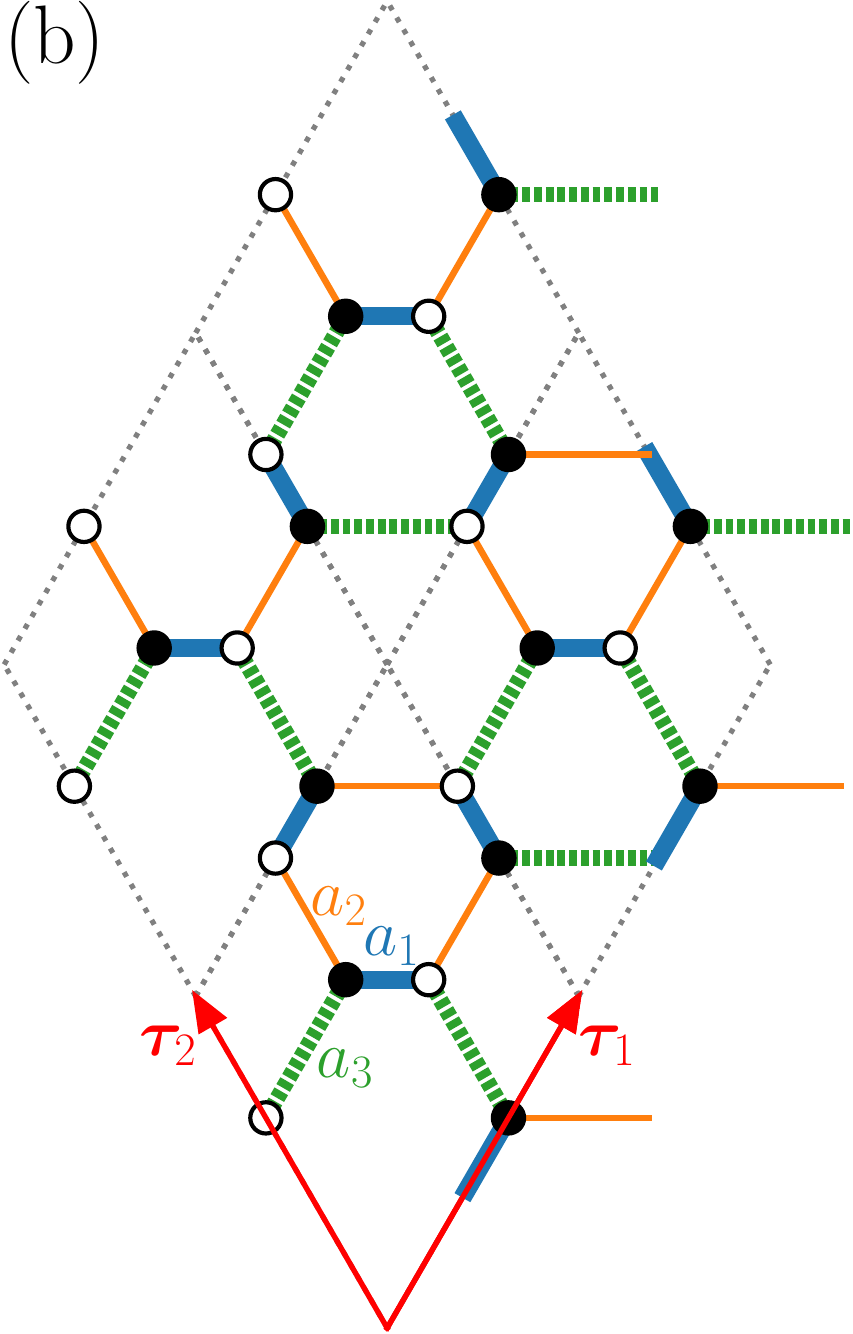}
  \caption{\label{fig:lattice}%
  (a) Undistorted ($a_{1} = a_{2} = a_{3} \equiv a$)
  and
  (b) maximally distorted ($a_{1} < a_{2} < a_{3}$)
  honeycomb lattices with the same volume (area) 
  preserving the reflection symmetry,
  where $a_{1}$, $a_{2}$, and $a_{3}$ denote lengths of
  the bonds depicted by heavy solid, light solid, and 
  dashed lines, respectively.
  Open and closed circles represent the two 
  sublattices, $A$ and $B$, 
  in the undistorted honeycomb lattice.
  With the possible distortions, each unit cell indicated
  by a trapezoid spanned by
  $\bm{\tau}_{1}=( \frac{3}{2}a, \frac{3\sqrt{3}}{2}a)$ and
  $\bm{\tau}_{2}=(-\frac{3}{2}a, \frac{3\sqrt{3}}{2}a)$ 
  contains six sites. 
  The periodic boundary conditions are imposed.
  }
 \end{figure}

In addition to the computational cost, there is another issue to consider
when using the QMC method in solving the self-consistent equation.
Since the results of the QMC simulation necessarily involve statistical errors,
the solution of the self-consistent equation also fluctuates.
To improve the quality of the solution within the limited computational resources,
we employ the Robbins-Monro algorithm, which is a simple and robust technique for 
root-finding problems from noisy observables~\cite{Robbins-AMS1951,Yasuda-PRE2013,Yasuda-PRB2015}.
We typically run about a hundred of the Robbins-Monro steps, 
in each of which a relatively short QMC simulation, e.g., 
1000 Monte Carlo steps for equilibration followed by 1000 steps for measurements
is performed on a finite-size cluster spanned by two lattice vectors
of $L \bm{\tau_{1}}$ and $L \bm{\tau_{2}}$. 
Here $\bm{\tau_{1}}$ and $\bm{\tau_{2}}$ are primitive translation vectors 
(see Fig.~\ref{fig:lattice})
of the unit cell for the distorted honeycomb lattice as described below.
The number of sites in the cluster is $N_{\mathrm{s}}= 6 L^2$, and 
we study the system with $L$ up to eight, 
which is moderate for the iterative calculations.

 \begin{figure}[tb]
  \centering
  \includegraphics[width=0.45\linewidth]{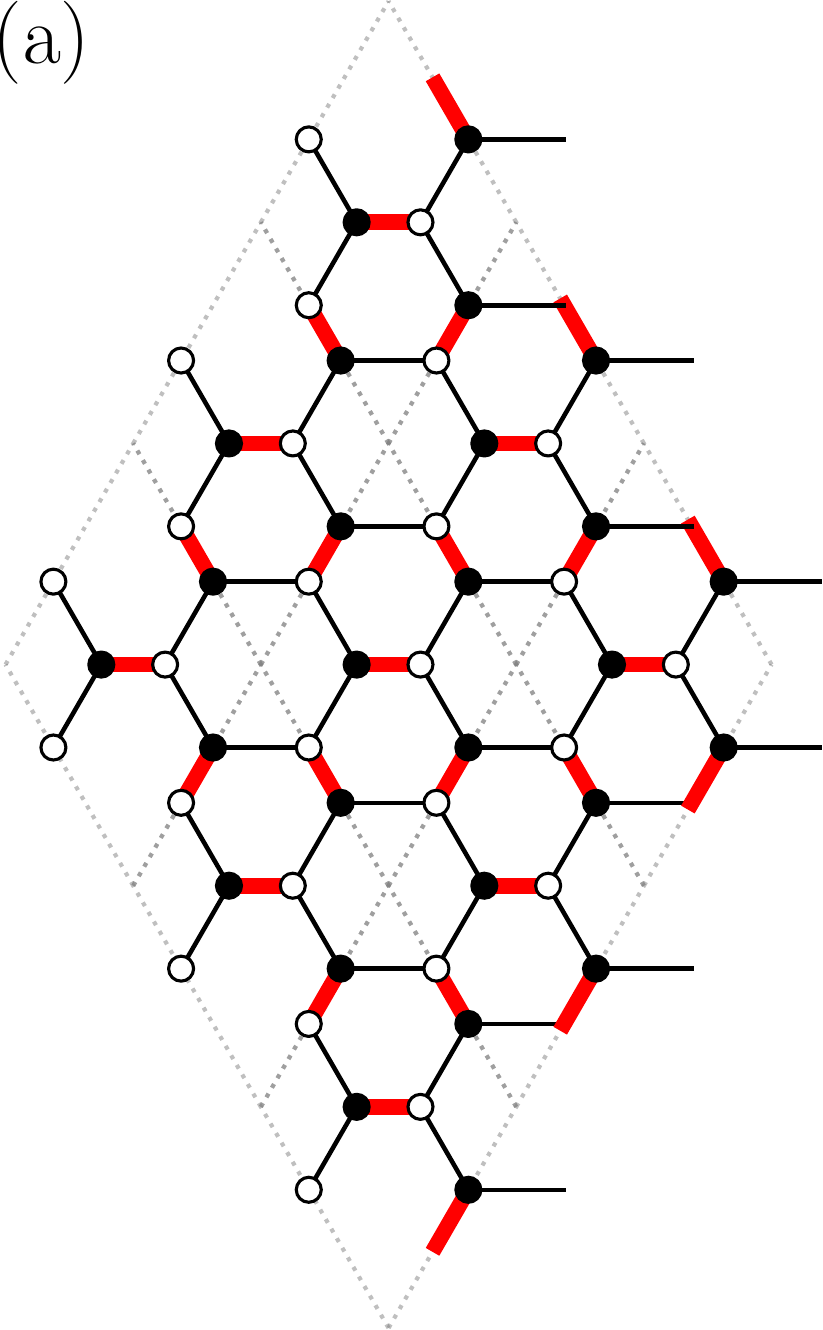}
  \includegraphics[width=0.45\linewidth]{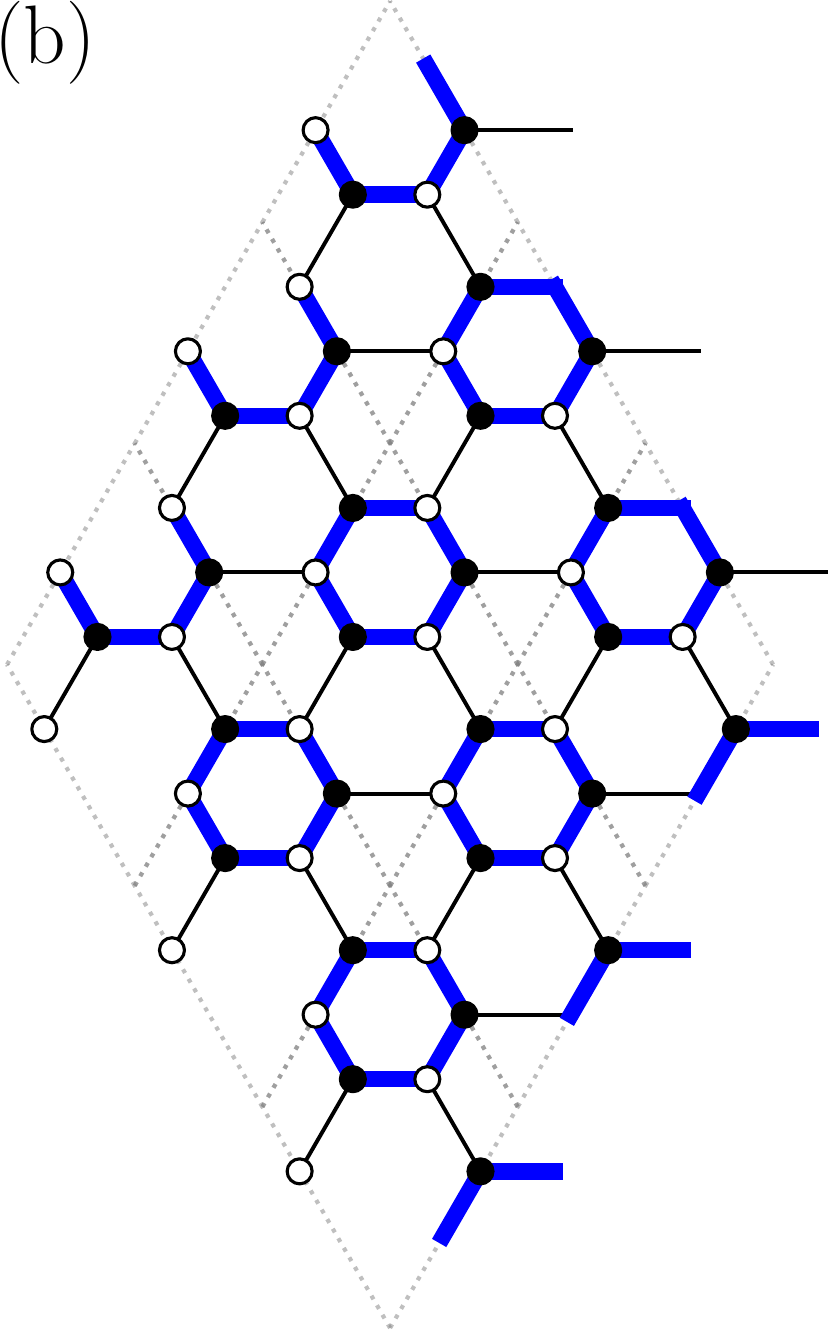}
  \caption{\label{fig:VBS}%
  Finite-size clusters ($L=3$) with
  (a) dimer-type distortion for the KVBS state
  and 
  (b) plaquette-type distortion for the KRVB state.
  In each figure, the shorter bonds are drawn with thick lines.
   }
 \end{figure}

We take the unit cell which contains six sites and three different bonds
as shown in Fig.~\ref{fig:lattice}.
We emphasize that
this is not an assumption of a supercell in a mean-field treatment.
Instead, it is what Frank and Lieb have rigorously shown based on the symmetry argument~\cite{Frank-PRL2011}.
Even if we take enlarged unit cells or consider all possible spatial
configurations without the unit cell, the most stable configuration should
converge to the above cell structure, possibly after a very long simulation time.
In Fig.~\ref{fig:lattice}, 
the bond lengths are denoted by $a_{i}=a\ ( 1 + u_{i} )$ ($i$=1,2,3),
and, without losing generality, we label these values as 
$u_{1} \le u_{2} \le u_{3}$.
Then, apart from the trivial undistorted lattice,
possible patterns of the lattice
distortions fall into three categories:
(1) maximum distortion        of $u_{1}<u_{2}<u_{3}$ [Fig.~\ref{fig:lattice}(b)],
(2) dimer-type distortion     of $u_{1}<u_{2}=u_{3}$ [Fig.~\ref{fig:VBS}(a)], and
(3) plaquette-type distortion of $u_{1}=u_{2}<u_{3}$ [Fig.~\ref{fig:VBS}(b)].
Here the state with the dimer-type distortion is a specific definition of
our KVBS state, while the state with the plaquette-type distortion is referred to as 
the Kekul\'{e} resonating-valence-bond (KRVB) state.
Furthermore, for simplicity, we study the system at a constant volume
by imposing a constraint condition $\sum_{i} u_{i}=0$.
This condition guarantees that the lengths of the primitive translation vectors 
are unchanged, i.e, $|\bm{\tau_{1 (2)}}| = \sum_{i} a_{i} = 3a$ (see Fig.~\ref{fig:lattice}),
thus preserving the volume of the system.
Subsequently, we can parameterize $u_{i}$ by two parameters, $u$ and $u^{\prime}$, as
$u_{1}=-u-u^{\prime}/2$, $u_{2}=u/2-u^{\prime}/2$, and $u_{3}=u/2+u^{\prime}$.
Solving the self-consistent equations, 
we obtain the optimized values of these parameters denoted as $\bar{u}$ and $\bar{u}^{\prime}$.
In this way,
the KVBS (KRVB) state is characterized by 
$\bar{u}>0$ and $\bar{u}^{\prime}=0$ ($\bar{u}=0$ and $\bar{u}^{\prime}>0$),
and the state with the maximum distortion is considered as a mixture of 
these two Kekul\'{e} states, i.e., $\bar{u}>0$ and $\bar{u}^{\prime}>0$.
We note that the constant volume condition is imposed just for simplicity. 
Since applying strain leads to an expansion of the volume, 
we could naively employ the lattice constant $a$ as a control parameter. 
In this case, the effect of the increasing $a$ is naturally incorporated 
in Eq.~(\ref{eq:hamiltonian}) by decreasing the transfer integral $t$, 
while the on-site Coulomb repulsion $U$ is expected to be less affected.
Then, the effect of strain can be mimicked by increasing $U/t$,
which is the conventional control parameter in the Hubbard model,
instead of actually increasing $a$.

\section{Results and Discussions}
\label{sec:results_and_discussions}

\begin{figure}[tb]
  \centering
  \includegraphics[width=1.0\linewidth]{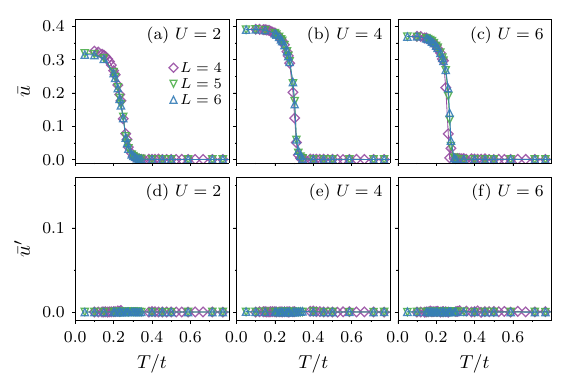}
  \caption{\label{fig:u-T-KI0060}%
  Temperature dependence of lattice displacements for $g/K=0.6$ with 
  (a, d) $U/t=2$, 
  (b, e) $U/t=4$, and
  (c, f) $U/t=6$.
  The upper and lower panels show 
  the order parameters of the KVBS and 
  the KRVB states, $\bar{u}$ and $\bar{u}^{\prime}$, respectively.
  }
\end{figure}

\begin{figure}[tb]
  \centering
  \includegraphics[width=1.0\linewidth]{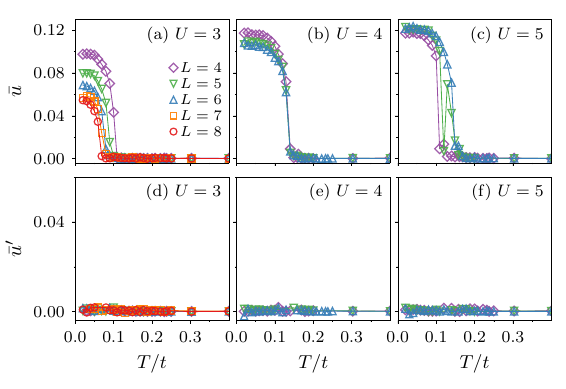}
  \caption{\label{fig:u-T-KI0040}%
  The same as Fig.~\ref{fig:u-T-KI0060} but for $g/K=0.4$ with
  (a, d) $U/t=3$, 
  (b, e) $U/t=4$, and
  (c, f) $U/t=5$. 
  }
\end{figure}

Let us begin by presenting the temperature dependence of the order parameters
for the two Kekul\'{e} states as shown in Figs.~\ref{fig:u-T-KI0060} 
and~\ref{fig:u-T-KI0040} for $g/K=0.6$ and $g/K=0.4$, respectively.
These values of the dimensionless parameter for the e-l coupling are 
representative for the soft and hard lattices in our model.
In the first place, out of the three possible lattice distortion patterns,
we observe that only the KVBS state is stabilized with $\bar{u}>0$ and $\bar{u}^{\prime}$=0 
for all parameters we studied at low temperatures.
We do not know why the KRVB state is not stabilized. 
Generally speaking, analytical methods would be eventually needed to 
understand the stability of different types of order~\cite{Roy-PRB2013,Roy-PRB2019}.
However, considering that the absence of the KRVB state is consistent with 
the \textit{ab initio} QMC study~\cite{Sorella-PRL2018}, 
we suppose that it is not simply due to the model simplification 
or the technical treatment, such as the constant volume condition. 
For the soft lattice, as shown in Fig.~\ref{fig:u-T-KI0060}, 
the phase transitions appear to be continuous 
for all $U/t$ and $L$, and the KVBS state is most stable with large $\bar{u}$ at 
$U/t=4 \simeq U_{\mathrm{c}}/t$.
On the other hand, in the case of the more rigid lattice  
(Fig.~\ref{fig:u-T-KI0040}), 
the KVBS state is fragile.
At $U/t=3$, the order parameter $\bar{u}$ depends on $L$,
converging to the thermodynamic limit value only with $L \ge 8$.
In addition, the converged value, i.e., $\bar{u} \simeq 0.055$, is quite small.
At $U/t=5$, the temperature dependence of $\bar{u}$ for the smaller lattice ($L <6$)
appears to signal that the transition is of first order.
However, this apparent behavior can be ascribed to the finite-size effect.
Since, for this large value of $U/t=5$, the antiferromagnetic (AF) correlation 
develops as $T/t$ decreases, its correlation length can reach the system 
size of small $L$ above the critical temperature of the Kekul\'{e} transition.
In this case, the transition to the KVBS state which is triggered by 
further lowering $T/t$ can share features with the transition between 
KVBS and AFMI at the ground state,
which is of first order as discussed later.
However, the result of $L=6$ in Fig.~\ref{fig:u-T-KI0040}(c) suggests that 
the Kekul\'{e} transition is more likely continuous.

\begin{figure}[tb]
  \centering
  \includegraphics[width=0.70\linewidth]{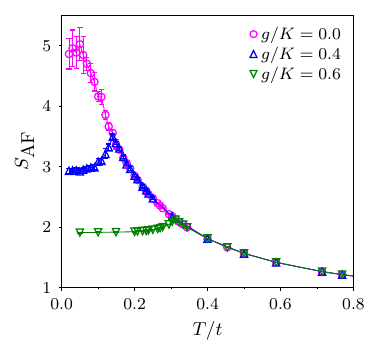}
  \caption{\label{fig:S_AF-T}%
 Temperature dependence of AF structure factor for 
 $g/K=0.0$, $0.4$, and $0.6$ with $U/t=4$ and $L=6$.
  }
\end{figure}

The competition between KVBS and AFMI is also observed in the magnetic property.
Figure~\ref{fig:S_AF-T} shows temperature dependence of the AF structure factor
defined as
\begin{equation}
 \label{eq:s_af}
  S_{\textrm{AF}} = \frac{1}{N_{\textrm{s}}}
  \left\langle
   \left(
    \sum_{i\in A} 
    \bm{S}_i
    -
    \sum_{i\in B} 
    \bm{S}_i
   \right)^{2}
  \right\rangle,
\end{equation}
where $\bm{S}_i=\frac{1}{2}\sum_{s, s^{\prime}}c^\dag_{is}(\bm{\sigma})_{ss'}c_{is'}$ is 
the spin operator at site $i$ with $\bm{\sigma}$ being the vector of Pauli matrices, 
and the sum $i\in A(B)$ runs over sites belonging to $A$ $(B)$ 
sublattices in the undistorted lattice [see Fig.~\ref{fig:lattice} (a)].
Since in the usual Hubbard model without the e-l coupling the AF long-range order
exists only in the ground state, $S_{\mathrm{AF}}$ for $g/K=0$ develops with decreasing
$T/t$ and saturates at a low temperature where the AF correlation length exceeds the 
system size $L$.
On the other hand, with the e-l coupling, $S_{\mathrm{AF}}$ shows a sudden drop 
at the KVBS transition temperature and does not develop thereafter,
which indicates that the KVBS state has no AF long-range order.

\begin{figure}[tb]
  \centering
  \includegraphics[width=0.90\linewidth]{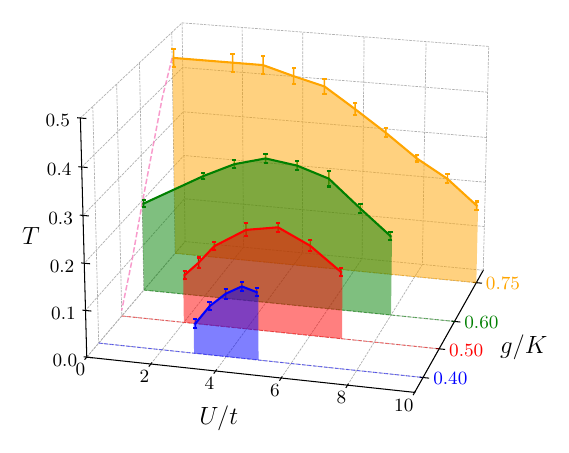}
  \caption{\label{fig:phase_diagram_3d}%
  Global phase diagram for the Kekul\'{e} VBS transition.
  The critical temperatures are plotted as a function of $U/t$ 
  for several values of $g/K$.
  Shaded areas represent regions of the Kekul\'{e} VBS phase.
  The dashed line in the $T/t$-$g/K$ plane at $U/t=0$ shows the 
  results of $L=16$.
  }
 \end{figure}

We determine the critical temperatures for the KVBS transition 
from the temperature-dependence of $\bar{u}$ for several values of $U/t$ and $g/K$,  
and the results are summarized in the finite-temperature phase diagram of Fig.~\ref{fig:phase_diagram_3d}.
The significant feature is that for small $g/K$, 
which corresponds to a more rigid lattice as in the case of graphene, 
the region of the KVBS phase is, although it is small, present in the vicinity 
of the critical point of the Hubbard model on the undistorted honeycomb lattice 
at $U_{\mathrm c}/t \simeq 3.8$%
~\cite{Sorella-SciRep2012,ParisenToldin-PRB2015,Otsuka-PRX2016,Seki-PRB2019,Ostmeyer-PRB2021}.
Thus, if the value of $U/t$ in real graphene is slightly smaller than $U_{\mathrm c}/t$,
which is indeed anticipated from the first-principles calculations of
$U=U_{00} = 9.3 \mathrm{eV}$~\cite{Wehling-PRL2011}
and $t \approx 2.7 \mathrm{eV}$~\cite{Reich-PRB2002,Jung-PRB2013}, 
application of strain is expected to bring about the phase transition to the KVBS state.

\begin{figure}[tb]
  \centering
  \includegraphics[width=1.0\linewidth]{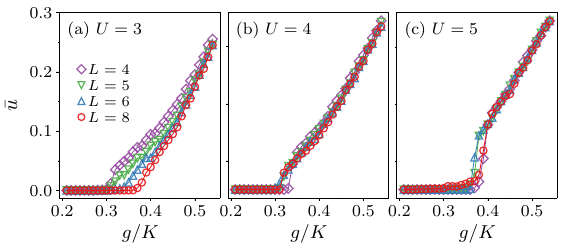}
  \caption{\label{fig:KI-umin-T0050}%
  Kekul\'{e} VBS order parameter $\bar{u}$ as a function of $g/K$ for 
  several system sizes $L$ with 
  (a) $U/t$ = 3, (b) $U/t$ = 4, and (c) $U/t$ = 5 at $T/t=0.05$ estimated by directly 
  evaluating $F(u)$.
  }
 \end{figure}

To investigate more closely the vicinity of $U_{\mathrm c}/t$
where the order parameter $\bar{u}$ is small, 
we directly calculate  the free energy as a function of $u$, i.e, $F(u)$ 
by numerically integrating its derivatives $\partial F/\partial u$,
which is given by the left-hand side of Eq.~(\ref{eq:self-consistent_equation}),
and determine $\bar{u}$ by locating the minimum of $F(u)$. 
This rather naive approach is computationally feasible since we have confirmed that 
the KRVB order does not appear and hence only one order parameter, $u$ for KVBS, is relevant in the free energy.
We calculate the derivatives at about a hundred discrete $u$ points 
by the QMC simulations, of which the computational cost is comparable to 
that of the iterative method using the Robbins-Monro algorithm.
The advantage of this method is that once the QMC simulations 
for the discrete $u$ points at a fixed value of $U/t$ are performed,
we can determine $F(u)$ and $\bar{u}$ for any value of $g/K$
as shown in Fig.~\ref{fig:KI-umin-T0050} 
for $U/t=3$, 4, and 5 at $T/t=0.05$.
For $U/t=3<U_{\mathrm{c}}/t$, the finite-size effect is evident even up to $L=8$.
This implies that a high resolution around the Dirac points in the 
momentum space is required to study the SM-KVBS transition at low temperatures.
On the other hand, for $U/t>U_{\mathrm{c}}/t$, where the ground-state of 
the system with $\bar{u}=0$ is AFMI, the finite-size effect
is small since the AF correlation length exceeds the system
size $L$, and the transition from AFMI to KVBS is confirmed to be of first order.

\begin{figure}[tb]
 \centering
 \includegraphics[width=0.9\linewidth]{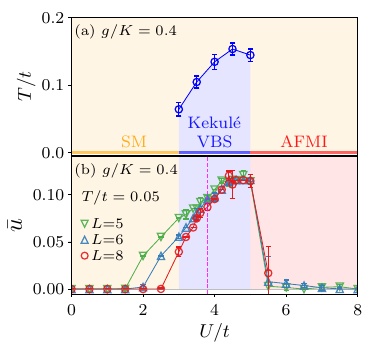}
 \caption{\label{fig:Tc-umin-KI0040}%
 $U/t$-dependence of 
 (a) the critical temperature of the Kekul\'{e} VBS transition for $g/K=0.4$
 and 
 (b) the Kekul\'{e} VBS order parameter $\bar{u}$ for the same $g/K=0.4$ at $T/t=0.05$.
 The dashed line in (b) indicates the location of the critical point
 separating SM and AFMI for the undistorted lattice, i.e.,
 $U_{\mathrm c}/t \simeq 3.8$.
 }
\end{figure}

We plot the KVBS order parameter $\bar{u}$ obtained by this approach at a fixed value of $g/K=0.4$
as a function of $U/t$ in Fig.~\ref{fig:Tc-umin-KI0040}(b), together with 
the critical temperature in Fig.~\ref{fig:Tc-umin-KI0040}(a) which is the
same data used in Fig.~\ref{fig:phase_diagram_3d}.
Now it is clear that the KVBS state emerges as an intermediate phase
between the SM and AFMI phases,
indicating that the Kekul\'{e}-like structural deformation is driven 
by the correlation before the AF long-range order sets in.
This result is primarily consistent with the preceding \textit{ab initio} 
QMC study~\cite{Sorella-PRL2018}.

\begin{figure}[tb]
 \centering
 \includegraphics[width=0.9\linewidth]{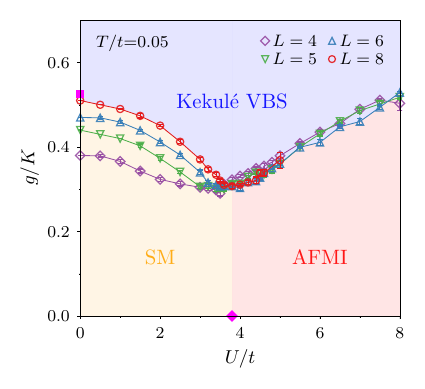}
 \caption{\label{fig:U-KIc-T0050}%
 Detailed phase diagram at $T/t=0.05$.
 Critical points of the phase boundary for each $L$ are determined
 by the $g/K$-dependence of $\bar{u}$ as shown in Fig.~\ref{fig:KI-umin-T0050}.
 For these moderate system sizes ($L \le 8$), the strong-coupling region
 without the Kekul\'{e} VBS order is effectively considered as AFMI at this
 low temperature, thus labeled as ``AFMI''.
 Solid diamond at $g/K=0$ indicates the critical point of the Hubbard model 
 on the undistorted honeycomb lattice, i.e., $U_{\mathrm c}/t \simeq 3.8$.
 Solid square at $U/t=0$ denotes the critical value of $g/K$ between 
 the Kekul\'{e} VBS and SM phases for a large lattice ($L=16$).
 }
\end{figure}

Finally, let us explore the phase diagram at a fixed temperature 
$T/t=0.05$ shown in Fig.~\ref{fig:U-KIc-T0050}.
As discussed above, the finite-size effect is strong in the phase boundary 
between the SM and KVBS phases. We calculate the 
critical value of $g/K$ for a large cluster $L=16$ in the noninteracting case 
and confirm that it is fairly close to that for $L=8$. 
Furthermore, there is less finite-size effect at $U/t \lesssim U_{\mathrm{c}}/t$. 
We thus expect that the phase boundary estimated for $L=8$ is sufficiently close 
to that in the thermodynamic limit.
On the other hand, the critical value of $g/K$ separating the KVBS and AFMI phases 
hardly depends on the system size.
This critical point is of first order and increases monotonically with $U$.
These features are in accord with recent results of the Su-Schrieffer-Heeger 
Hubbard (SSHH) model on the square lattice, where quantum phonon dynamics is fully 
taken into account~\cite{Cai-PRL2021,Cai-PRB2022,Feng-PRB2022}.
The first-order transition between the VBS and AF phases is also found in
the spin-Peierls model on the honeycomb lattice~\cite{Weber-PRB2021}, which
corresponds to the SSHH model in the strong coupling limit ($U/t>>1$).
Therefore, we expect that the first-order nature of the transition between
these phases, being within the Landau-Ginzburg-Wilson paradigm, is ubiquitous 
especially near the adiabatic limit~\cite{Weber-PRB2021}.
It is also pointed out that our phase diagram featuring a tricritical point
with the SM, AFMI, and KVBS phases is similar to that obtained for a model
of Dirac fermions devised from a designer Hamiltonian approach~\cite{Sato-PRL2017},
except that the AFMI-KVBS transition is continuous because of the emergent 
SO(4) symmetry in the devised model.

\section{Summary}
\label{sec:summary}

To summarize, we have investigated the valence-bond-solid phase in the Hubbard 
model on the honeycomb lattice with the electron-lattice coupling,
motivated by the recent \textit{ab initio} calculation predicting
the correlation-driven Kekul\'{e} valence-bond-solid phase in graphene.
Considering the lattice displacements 
in the adiabatic limit, the model is readily trackable by 
the auxiliary-field quantum Monte Carlo method.
Furthermore, based on the symmetry argument by Frank and Lieb,
the possible distortions in the Hubbard model for graphene are
restricted to have the unit cell containing at most six sites.
Under this restriction as well as the constant-volume condition,
we have obtained the finite-temperature phase diagram by determining
the most stable displacement configuration.
Out of the three possible bond-order patterns,
only the Kekul\'{e} valence-bond-solid phase with the dimerized distortion 
is found to be stabilized. 
For a more rigid lattice, as in the case of graphene, the Kekul\'{e}
valence-bond-solid phase exists in a narrow region near the critical 
point of the metal-insulator transition in the Hubbard model without 
the electron-lattice coupling.
This implies that if the real graphene is located in the semimetallic
side near this critical point in the phase diagram as a function of the
interaction, applying strain, which should correspond to increasing the effective interaction, 
can bring about the valence-bond-order transition and hence introduce the gap in graphene.
We have also discussed the phase diagram at a low temperature, where 
the semimetal, the antiferomagnetic Mott insulator, and the Kekul\'{e}
valence-bond-solid phases are competing. 
The phase boundary between the antiferromagnetic Mott insulator and the Kekul\'{e} phase
is confirmed to be of the first order, which falls in the Landau-Ginzburg-Wilson paradigm.

\acknowledgments
This work was supported by JSPS KAKENHI Grant Numbers 
JP18K03475, JP21K03395, and JP21H04446.
It is also partially supported by Program for Promoting Research on the Supercomputer 
Fugaku (No.~JPMXP1020230411) from MEXT, Japan, and by the COE research grant 
in computational science from Hyogo Prefecture and Kobe City through Foundation for Computational Science.
Parts of numerical simulations have been performed on 
the HOKUSAI supercomputer at RIKEN (Project IDs: Q22525 and Q23609) and 
the FUGAKU supercomputer provided by the RIKEN Center for Computational Science (R-CCS)
through the HPCI System Research project (Project ID: hp220138). 
This paper is written in memory of Sandro Sorella, who sadly passed away in August 2022.

\bibliography{distorted}

\begin{thebibliography}{70}%
\makeatletter
\providecommand \@ifxundefined [1]{%
 \@ifx{#1\undefined}
}%
\providecommand \@ifnum [1]{%
 \ifnum #1\expandafter \@firstoftwo
 \else \expandafter \@secondoftwo
 \fi
}%
\providecommand \@ifx [1]{%
 \ifx #1\expandafter \@firstoftwo
 \else \expandafter \@secondoftwo
 \fi
}%
\providecommand \natexlab [1]{#1}%
\providecommand \enquote  [1]{``#1''}%
\providecommand \bibnamefont  [1]{#1}%
\providecommand \bibfnamefont [1]{#1}%
\providecommand \citenamefont [1]{#1}%
\providecommand \href@noop [0]{\@secondoftwo}%
\providecommand \href [0]{\begingroup \@sanitize@url \@href}%
\providecommand \@href[1]{\@@startlink{#1}\@@href}%
\providecommand \@@href[1]{\endgroup#1\@@endlink}%
\providecommand \@sanitize@url [0]{\catcode `\\12\catcode `\$12\catcode
  `\&12\catcode `\#12\catcode `\^12\catcode `\_12\catcode `\%12\relax}%
\providecommand \@@startlink[1]{}%
\providecommand \@@endlink[0]{}%
\providecommand \url  [0]{\begingroup\@sanitize@url \@url }%
\providecommand \@url [1]{\endgroup\@href {#1}{\urlprefix }}%
\providecommand \urlprefix  [0]{URL }%
\providecommand \Eprint [0]{\href }%
\providecommand \doibase [0]{https://doi.org/}%
\providecommand \selectlanguage [0]{\@gobble}%
\providecommand \bibinfo  [0]{\@secondoftwo}%
\providecommand \bibfield  [0]{\@secondoftwo}%
\providecommand \translation [1]{[#1]}%
\providecommand \BibitemOpen [0]{}%
\providecommand \bibitemStop [0]{}%
\providecommand \bibitemNoStop [0]{.\EOS\space}%
\providecommand \EOS [0]{\spacefactor3000\relax}%
\providecommand \BibitemShut  [1]{\csname bibitem#1\endcsname}%
\let\auto@bib@innerbib\@empty
\bibitem [{\citenamefont {Sorella}\ and\ \citenamefont
  {Tosatti}(1992)}]{Sorella-EPL1992}%
  \BibitemOpen
  \bibfield  {author} {\bibinfo {author} {\bibfnamefont {S.}~\bibnamefont
  {Sorella}}\ and\ \bibinfo {author} {\bibfnamefont {E.}~\bibnamefont
  {Tosatti}},\ }\bibfield  {title} {\bibinfo {title} {Semi-{{Metal-Insulator
  Transition}} of the {{Hubbard Model}} in the {{Honeycomb Lattice}}},\ }\href
  {https://doi.org/10.1209/0295-5075/19/8/007} {\bibfield  {journal} {\bibinfo
  {journal} {Europhys. Lett.}\ }\textbf {\bibinfo {volume} {19}},\ \bibinfo
  {pages} {699} (\bibinfo {year} {1992})}\BibitemShut {NoStop}%
\bibitem [{\citenamefont {Novoselov}\ \emph {et~al.}(2004)\citenamefont
  {Novoselov}, \citenamefont {Geim}, \citenamefont {Morozov}, \citenamefont
  {Jiang}, \citenamefont {Zhang}, \citenamefont {Dubonos}, \citenamefont
  {Grigorieva},\ and\ \citenamefont {Firsov}}]{Novoselov-Science2004}%
  \BibitemOpen
  \bibfield  {author} {\bibinfo {author} {\bibfnamefont {K.~S.}\ \bibnamefont
  {Novoselov}}, \bibinfo {author} {\bibfnamefont {A.~K.}\ \bibnamefont {Geim}},
  \bibinfo {author} {\bibfnamefont {S.~V.}\ \bibnamefont {Morozov}}, \bibinfo
  {author} {\bibfnamefont {D.}~\bibnamefont {Jiang}}, \bibinfo {author}
  {\bibfnamefont {Y.}~\bibnamefont {Zhang}}, \bibinfo {author} {\bibfnamefont
  {S.~V.}\ \bibnamefont {Dubonos}}, \bibinfo {author} {\bibfnamefont {I.~V.}\
  \bibnamefont {Grigorieva}},\ and\ \bibinfo {author} {\bibfnamefont {A.~A.}\
  \bibnamefont {Firsov}},\ }\bibfield  {title} {\bibinfo {title} {Electric
  {{Field Effect}} in {{Atomically Thin Carbon Films}}},\ }\href
  {https://doi.org/10.1126/science.1102896} {\bibfield  {journal} {\bibinfo
  {journal} {Science}\ }\textbf {\bibinfo {volume} {306}},\ \bibinfo {pages}
  {666} (\bibinfo {year} {2004})}\BibitemShut {NoStop}%
\bibitem [{\citenamefont {Castro~Neto}\ \emph {et~al.}(2009)\citenamefont
  {Castro~Neto}, \citenamefont {Guinea}, \citenamefont {Peres}, \citenamefont
  {Novoselov},\ and\ \citenamefont {Geim}}]{CastroNeto-RMP2009}%
  \BibitemOpen
  \bibfield  {author} {\bibinfo {author} {\bibfnamefont {A.~H.}\ \bibnamefont
  {Castro~Neto}}, \bibinfo {author} {\bibfnamefont {F.}~\bibnamefont {Guinea}},
  \bibinfo {author} {\bibfnamefont {N.~M.~R.}\ \bibnamefont {Peres}}, \bibinfo
  {author} {\bibfnamefont {K.~S.}\ \bibnamefont {Novoselov}},\ and\ \bibinfo
  {author} {\bibfnamefont {A.~K.}\ \bibnamefont {Geim}},\ }\bibfield  {title}
  {\bibinfo {title} {The electronic properties of graphene},\ }\href
  {https://doi.org/10.1103/RevModPhys.81.109} {\bibfield  {journal} {\bibinfo
  {journal} {Rev. Mod. Phys.}\ }\textbf {\bibinfo {volume} {81}},\ \bibinfo
  {pages} {109} (\bibinfo {year} {2009})}\BibitemShut {NoStop}%
\bibitem [{\citenamefont {Das~Sarma}\ \emph {et~al.}(2011)\citenamefont
  {Das~Sarma}, \citenamefont {Adam}, \citenamefont {Hwang},\ and\ \citenamefont
  {Rossi}}]{DasSarma-RMP2011}%
  \BibitemOpen
  \bibfield  {author} {\bibinfo {author} {\bibfnamefont {S.}~\bibnamefont
  {Das~Sarma}}, \bibinfo {author} {\bibfnamefont {S.}~\bibnamefont {Adam}},
  \bibinfo {author} {\bibfnamefont {E.~H.}\ \bibnamefont {Hwang}},\ and\
  \bibinfo {author} {\bibfnamefont {E.}~\bibnamefont {Rossi}},\ }\bibfield
  {title} {\bibinfo {title} {Electronic transport in two-dimensional
  graphene},\ }\href {https://doi.org/10.1103/RevModPhys.83.407} {\bibfield
  {journal} {\bibinfo  {journal} {Rev. Mod. Phys.}\ }\textbf {\bibinfo {volume}
  {83}},\ \bibinfo {pages} {407} (\bibinfo {year} {2011})}\BibitemShut
  {NoStop}%
\bibitem [{\citenamefont {Kotov}\ \emph {et~al.}(2012)\citenamefont {Kotov},
  \citenamefont {Uchoa}, \citenamefont {Pereira}, \citenamefont {Guinea},\ and\
  \citenamefont {Castro~Neto}}]{Kotov-RMP2012}%
  \BibitemOpen
  \bibfield  {author} {\bibinfo {author} {\bibfnamefont {V.~N.}\ \bibnamefont
  {Kotov}}, \bibinfo {author} {\bibfnamefont {B.}~\bibnamefont {Uchoa}},
  \bibinfo {author} {\bibfnamefont {V.~M.}\ \bibnamefont {Pereira}}, \bibinfo
  {author} {\bibfnamefont {F.}~\bibnamefont {Guinea}},\ and\ \bibinfo {author}
  {\bibfnamefont {A.~H.}\ \bibnamefont {Castro~Neto}},\ }\bibfield  {title}
  {\bibinfo {title} {Electron-electron interactions in graphene: {{Current}}
  status and perspectives},\ }\href
  {https://doi.org/10.1103/RevModPhys.84.1067} {\bibfield  {journal} {\bibinfo
  {journal} {Rev. Mod. Phys.}\ }\textbf {\bibinfo {volume} {84}},\ \bibinfo
  {pages} {1067} (\bibinfo {year} {2012})}\BibitemShut {NoStop}%
\bibitem [{\citenamefont {Vafek}\ and\ \citenamefont
  {Vishwanath}(2014)}]{Vafek-Review2014}%
  \BibitemOpen
  \bibfield  {author} {\bibinfo {author} {\bibfnamefont {O.}~\bibnamefont
  {Vafek}}\ and\ \bibinfo {author} {\bibfnamefont {A.}~\bibnamefont
  {Vishwanath}},\ }\bibfield  {title} {\bibinfo {title} {Dirac fermions in
  solids: From high-{{$T_{\mathrm{c}}$}} cuprates and graphene to topological
  insulators and weyl semimetals},\ }\href
  {https://doi.org/10.1146/annurev-conmatphys-031113-133841} {\bibfield
  {journal} {\bibinfo  {journal} {Annu. Rev. Condens. Matter Phys.}\ }\textbf
  {\bibinfo {volume} {5}},\ \bibinfo {pages} {83} (\bibinfo {year}
  {2014})}\BibitemShut {NoStop}%
\bibitem [{\citenamefont {Wehling}\ \emph {et~al.}(2014)\citenamefont
  {Wehling}, \citenamefont {{Black-Schaffer}},\ and\ \citenamefont
  {Balatsky}}]{Wehling-Review2014}%
  \BibitemOpen
  \bibfield  {author} {\bibinfo {author} {\bibfnamefont {T.}~\bibnamefont
  {Wehling}}, \bibinfo {author} {\bibfnamefont {A.}~\bibnamefont
  {{Black-Schaffer}}},\ and\ \bibinfo {author} {\bibfnamefont {A.}~\bibnamefont
  {Balatsky}},\ }\bibfield  {title} {\bibinfo {title} {Dirac materials},\
  }\href {https://doi.org/10.1080/00018732.2014.927109} {\bibfield  {journal}
  {\bibinfo  {journal} {Adv. Phys.}\ }\textbf {\bibinfo {volume} {63}},\
  \bibinfo {pages} {1} (\bibinfo {year} {2014})}\BibitemShut {NoStop}%
\bibitem [{\citenamefont {Boyack}\ \emph {et~al.}(2021)\citenamefont {Boyack},
  \citenamefont {Yerzhakov},\ and\ \citenamefont {Maciejko}}]{Boyack-EPJ2021}%
  \BibitemOpen
  \bibfield  {author} {\bibinfo {author} {\bibfnamefont {R.}~\bibnamefont
  {Boyack}}, \bibinfo {author} {\bibfnamefont {H.}~\bibnamefont {Yerzhakov}},\
  and\ \bibinfo {author} {\bibfnamefont {J.}~\bibnamefont {Maciejko}},\
  }\bibfield  {title} {\bibinfo {title} {Quantum phase transitions in {{Dirac}}
  fermion systems},\ }\href {https://doi.org/10.1140/epjs/s11734-021-00069-1}
  {\bibfield  {journal} {\bibinfo  {journal} {Eur. Phys. J. Spec. Top.}\
  }\textbf {\bibinfo {volume} {230}},\ \bibinfo {pages} {979} (\bibinfo {year}
  {2021})}\BibitemShut {NoStop}%
\bibitem [{\citenamefont {Smith}\ \emph {et~al.}(2021)\citenamefont {Smith},
  \citenamefont {Buividovich}, \citenamefont {K{\"o}rner}, \citenamefont
  {Ulybyshev},\ and\ \citenamefont {{von Smekal}}}]{Smith-IJMPA2021}%
  \BibitemOpen
  \bibfield  {author} {\bibinfo {author} {\bibfnamefont {D.}~\bibnamefont
  {Smith}}, \bibinfo {author} {\bibfnamefont {P.}~\bibnamefont {Buividovich}},
  \bibinfo {author} {\bibfnamefont {M.}~\bibnamefont {K{\"o}rner}}, \bibinfo
  {author} {\bibfnamefont {M.}~\bibnamefont {Ulybyshev}},\ and\ \bibinfo
  {author} {\bibfnamefont {L.}~\bibnamefont {{von Smekal}}},\ }\bibfield
  {title} {\bibinfo {title} {Quantum phase transitions on the hexagonal
  lattice},\ }\href {https://doi.org/10.1142/S0217751X21410037} {\bibfield
  {journal} {\bibinfo  {journal} {Int. J. Mod. Phys. A}\ }\textbf {\bibinfo
  {volume} {36}},\ \bibinfo {pages} {2141003} (\bibinfo {year}
  {2021})}\BibitemShut {NoStop}%
\bibitem [{\citenamefont {Meng}\ \emph {et~al.}(2010)\citenamefont {Meng},
  \citenamefont {Lang}, \citenamefont {Wessel}, \citenamefont {Assaad},\ and\
  \citenamefont {Muramatsu}}]{Meng-Nature2010}%
  \BibitemOpen
  \bibfield  {author} {\bibinfo {author} {\bibfnamefont {Z.~Y.}\ \bibnamefont
  {Meng}}, \bibinfo {author} {\bibfnamefont {T.~C.}\ \bibnamefont {Lang}},
  \bibinfo {author} {\bibfnamefont {S.}~\bibnamefont {Wessel}}, \bibinfo
  {author} {\bibfnamefont {F.~F.}\ \bibnamefont {Assaad}},\ and\ \bibinfo
  {author} {\bibfnamefont {A.}~\bibnamefont {Muramatsu}},\ }\bibfield  {title}
  {\bibinfo {title} {Quantum spin liquid emerging in two-dimensional correlated
  {{Dirac}} fermions.},\ }\href {https://doi.org/10.1038/nature08942}
  {\bibfield  {journal} {\bibinfo  {journal} {Nature}\ }\textbf {\bibinfo
  {volume} {464}},\ \bibinfo {pages} {847} (\bibinfo {year}
  {2010})}\BibitemShut {NoStop}%
\bibitem [{\citenamefont {Chang}\ and\ \citenamefont
  {Scalettar}(2012)}]{Chang-PRL2012}%
  \BibitemOpen
  \bibfield  {author} {\bibinfo {author} {\bibfnamefont {C.-C.}\ \bibnamefont
  {Chang}}\ and\ \bibinfo {author} {\bibfnamefont {R.~T.}\ \bibnamefont
  {Scalettar}},\ }\bibfield  {title} {\bibinfo {title} {Quantum {{Disordered
  Phase}} near the {{Mott Transition}} in the {{Staggered-Flux Hubbard Model}}
  on a {{Square Lattice}}},\ }\href
  {https://doi.org/10.1103/PhysRevLett.109.026404} {\bibfield  {journal}
  {\bibinfo  {journal} {Phys. Rev. Lett.}\ }\textbf {\bibinfo {volume} {109}},\
  \bibinfo {pages} {026404} (\bibinfo {year} {2012})}\BibitemShut {NoStop}%
\bibitem [{\citenamefont {Seki}\ and\ \citenamefont {Ohta}()}]{Seki-arXiv2012}%
  \BibitemOpen
  \bibfield  {author} {\bibinfo {author} {\bibfnamefont {K.}~\bibnamefont
  {Seki}}\ and\ \bibinfo {author} {\bibfnamefont {Y.}~\bibnamefont {Ohta}},\
  }\href {http://arxiv.org/abs/1209.2101} {\bibinfo {title} {Quantum phase
  transitions in the honeycomb-lattice {{Hubbard}} model}},\ \Eprint
  {https://arxiv.org/abs/1209.2101} {arXiv:1209.2101} \BibitemShut {NoStop}%
\bibitem [{\citenamefont {Sorella}\ \emph {et~al.}(2012)\citenamefont
  {Sorella}, \citenamefont {Otsuka},\ and\ \citenamefont
  {Yunoki}}]{Sorella-SciRep2012}%
  \BibitemOpen
  \bibfield  {author} {\bibinfo {author} {\bibfnamefont {S.}~\bibnamefont
  {Sorella}}, \bibinfo {author} {\bibfnamefont {Y.}~\bibnamefont {Otsuka}},\
  and\ \bibinfo {author} {\bibfnamefont {S.}~\bibnamefont {Yunoki}},\
  }\bibfield  {title} {\bibinfo {title} {2},\ }\href
  {https://doi.org/10.1038/srep00992} {\bibfield  {journal} {\bibinfo
  {journal} {Sci. Rep.}\ }\textbf {\bibinfo {volume} {2}},\ \bibinfo {pages}
  {992} (\bibinfo {year} {2012})}\BibitemShut {NoStop}%
\bibitem [{\citenamefont {Hassan}\ and\ \citenamefont
  {S{\'e}n{\'e}chal}(2013)}]{Hassan-PRL2013}%
  \BibitemOpen
  \bibfield  {author} {\bibinfo {author} {\bibfnamefont {S.~R.}\ \bibnamefont
  {Hassan}}\ and\ \bibinfo {author} {\bibfnamefont {D.}~\bibnamefont
  {S{\'e}n{\'e}chal}},\ }\bibfield  {title} {\bibinfo {title} {Absence of
  {{Spin Liquid}} in {{Nonfrustrated Correlated Systems}}},\ }\href
  {https://doi.org/10.1103/PhysRevLett.110.096402} {\bibfield  {journal}
  {\bibinfo  {journal} {Phys. Rev. Lett.}\ }\textbf {\bibinfo {volume} {110}},\
  \bibinfo {pages} {096402} (\bibinfo {year} {2013})}\BibitemShut {NoStop}%
\bibitem [{\citenamefont {Chen}\ \emph {et~al.}(2014)\citenamefont {Chen},
  \citenamefont {Booth}, \citenamefont {Sharma}, \citenamefont {Knizia},\ and\
  \citenamefont {Chan}}]{Chen-PRB2014}%
  \BibitemOpen
  \bibfield  {author} {\bibinfo {author} {\bibfnamefont {Q.}~\bibnamefont
  {Chen}}, \bibinfo {author} {\bibfnamefont {G.~H.}\ \bibnamefont {Booth}},
  \bibinfo {author} {\bibfnamefont {S.}~\bibnamefont {Sharma}}, \bibinfo
  {author} {\bibfnamefont {G.}~\bibnamefont {Knizia}},\ and\ \bibinfo {author}
  {\bibfnamefont {G.~K.-L.}\ \bibnamefont {Chan}},\ }\bibfield  {title}
  {\bibinfo {title} {Intermediate and spin-liquid phase of the half-filled
  honeycomb {{Hubbard}} model},\ }\href
  {https://doi.org/10.1103/PhysRevB.89.165134} {\bibfield  {journal} {\bibinfo
  {journal} {Phys. Rev. B}\ }\textbf {\bibinfo {volume} {89}},\ \bibinfo
  {pages} {165134} (\bibinfo {year} {2014})}\BibitemShut {NoStop}%
\bibitem [{\citenamefont {Ixert}\ \emph {et~al.}(2014)\citenamefont {Ixert},
  \citenamefont {Assaad},\ and\ \citenamefont {Schmidt}}]{Ixert-PRB2014}%
  \BibitemOpen
  \bibfield  {author} {\bibinfo {author} {\bibfnamefont {D.}~\bibnamefont
  {Ixert}}, \bibinfo {author} {\bibfnamefont {F.~F.}\ \bibnamefont {Assaad}},\
  and\ \bibinfo {author} {\bibfnamefont {K.~P.}\ \bibnamefont {Schmidt}},\
  }\bibfield  {title} {\bibinfo {title} {Mott physics in the half-filled
  {{Hubbard}} model on a family of vortex-full square lattices},\ }\href
  {https://doi.org/10.1103/PhysRevB.90.195133} {\bibfield  {journal} {\bibinfo
  {journal} {Phys. Rev. B}\ }\textbf {\bibinfo {volume} {90}},\ \bibinfo
  {pages} {195133} (\bibinfo {year} {2014})}\BibitemShut {NoStop}%
\bibitem [{\citenamefont {Herbut}(2006)}]{Herbut-PRL2006}%
  \BibitemOpen
  \bibfield  {author} {\bibinfo {author} {\bibfnamefont {I.~F.}\ \bibnamefont
  {Herbut}},\ }\bibfield  {title} {\bibinfo {title} {Interactions and {{Phase
  Transitions}} on {{Graphene}}'s {{Honeycomb Lattice}}},\ }\href
  {https://doi.org/10.1103/PhysRevLett.97.146401} {\bibfield  {journal}
  {\bibinfo  {journal} {Phys. Rev. Lett.}\ }\textbf {\bibinfo {volume} {97}},\
  \bibinfo {pages} {146401} (\bibinfo {year} {2006})}\BibitemShut {NoStop}%
\bibitem [{\citenamefont {Herbut}\ \emph
  {et~al.}(2009{\natexlab{a}})\citenamefont {Herbut}, \citenamefont {Juri{\v
  c}i{\'c}},\ and\ \citenamefont {Roy}}]{Herbut-PRB2009a}%
  \BibitemOpen
  \bibfield  {author} {\bibinfo {author} {\bibfnamefont {I.~F.}\ \bibnamefont
  {Herbut}}, \bibinfo {author} {\bibfnamefont {V.}~\bibnamefont {Juri{\v
  c}i{\'c}}},\ and\ \bibinfo {author} {\bibfnamefont {B.}~\bibnamefont {Roy}},\
  }\bibfield  {title} {\bibinfo {title} {Theory of interacting electrons on the
  honeycomb lattice},\ }\href {https://doi.org/10.1103/PhysRevB.79.085116}
  {\bibfield  {journal} {\bibinfo  {journal} {Phys. Rev. B}\ }\textbf {\bibinfo
  {volume} {79}},\ \bibinfo {pages} {085116} (\bibinfo {year}
  {2009}{\natexlab{a}})}\BibitemShut {NoStop}%
\bibitem [{\citenamefont {Herbut}\ \emph
  {et~al.}(2009{\natexlab{b}})\citenamefont {Herbut}, \citenamefont {Juri{\v
  c}i{\'c}},\ and\ \citenamefont {Vafek}}]{Herbut-PRB2009b}%
  \BibitemOpen
  \bibfield  {author} {\bibinfo {author} {\bibfnamefont {I.~F.}\ \bibnamefont
  {Herbut}}, \bibinfo {author} {\bibfnamefont {V.}~\bibnamefont {Juri{\v
  c}i{\'c}}},\ and\ \bibinfo {author} {\bibfnamefont {O.}~\bibnamefont
  {Vafek}},\ }\bibfield  {title} {\bibinfo {title} {Relativistic {{Mott}}
  criticality in graphene},\ }\href
  {https://doi.org/10.1103/PhysRevB.80.075432} {\bibfield  {journal} {\bibinfo
  {journal} {Phys. Rev. B}\ }\textbf {\bibinfo {volume} {80}},\ \bibinfo
  {pages} {075432} (\bibinfo {year} {2009}{\natexlab{b}})}\BibitemShut
  {NoStop}%
\bibitem [{\citenamefont {Ryu}\ \emph {et~al.}(2009)\citenamefont {Ryu},
  \citenamefont {Mudry}, \citenamefont {Hou},\ and\ \citenamefont
  {Chamon}}]{Ryu-PRB2009}%
  \BibitemOpen
  \bibfield  {author} {\bibinfo {author} {\bibfnamefont {S.}~\bibnamefont
  {Ryu}}, \bibinfo {author} {\bibfnamefont {C.}~\bibnamefont {Mudry}}, \bibinfo
  {author} {\bibfnamefont {C.-Y.}\ \bibnamefont {Hou}},\ and\ \bibinfo {author}
  {\bibfnamefont {C.}~\bibnamefont {Chamon}},\ }\bibfield  {title} {\bibinfo
  {title} {Masses in graphenelike two-dimensional electronic systems:
  {{Topological}} defects in order parameters and their fractional exchange
  statistics},\ }\href {https://doi.org/10.1103/PhysRevB.80.205319} {\bibfield
  {journal} {\bibinfo  {journal} {Phys. Rev. B}\ }\textbf {\bibinfo {volume}
  {80}},\ \bibinfo {pages} {205319} (\bibinfo {year} {2009})}\BibitemShut
  {NoStop}%
\bibitem [{\citenamefont {Assaad}\ and\ \citenamefont
  {Herbut}(2013)}]{Assaad-PRX2013}%
  \BibitemOpen
  \bibfield  {author} {\bibinfo {author} {\bibfnamefont {F.~F.}\ \bibnamefont
  {Assaad}}\ and\ \bibinfo {author} {\bibfnamefont {I.~F.}\ \bibnamefont
  {Herbut}},\ }\bibfield  {title} {\bibinfo {title} {Pinning the order: The
  nature of quantum criticality in the {{Hubbard}} model on honeycomb
  lattice},\ }\href {https://doi.org/10.1103/PhysRevX.3.031010} {\bibfield
  {journal} {\bibinfo  {journal} {Phys. Rev. X}\ }\textbf {\bibinfo {volume}
  {3}},\ \bibinfo {pages} {031010} (\bibinfo {year} {2013})}\BibitemShut
  {NoStop}%
\bibitem [{\citenamefont {Parisen~Toldin}\ \emph {et~al.}(2015)\citenamefont
  {Parisen~Toldin}, \citenamefont {Hohenadler}, \citenamefont {Assaad},\ and\
  \citenamefont {Herbut}}]{ParisenToldin-PRB2015}%
  \BibitemOpen
  \bibfield  {author} {\bibinfo {author} {\bibfnamefont {F.}~\bibnamefont
  {Parisen~Toldin}}, \bibinfo {author} {\bibfnamefont {M.}~\bibnamefont
  {Hohenadler}}, \bibinfo {author} {\bibfnamefont {F.~F.}\ \bibnamefont
  {Assaad}},\ and\ \bibinfo {author} {\bibfnamefont {I.~F.}\ \bibnamefont
  {Herbut}},\ }\bibfield  {title} {\bibinfo {title} {Fermionic quantum
  criticality in honeycomb and {$\pi$}-flux hubbard models: Finite-size scaling
  of renormalization-group-invariant observables from quantum monte carlo},\
  }\href {https://doi.org/10.1103/PhysRevB.91.165108} {\bibfield  {journal}
  {\bibinfo  {journal} {Phys. Rev. B}\ }\textbf {\bibinfo {volume} {91}},\
  \bibinfo {pages} {165108} (\bibinfo {year} {2015})}\BibitemShut {NoStop}%
\bibitem [{\citenamefont {Otsuka}\ \emph {et~al.}(2016)\citenamefont {Otsuka},
  \citenamefont {Yunoki},\ and\ \citenamefont {Sorella}}]{Otsuka-PRX2016}%
  \BibitemOpen
  \bibfield  {author} {\bibinfo {author} {\bibfnamefont {Y.}~\bibnamefont
  {Otsuka}}, \bibinfo {author} {\bibfnamefont {S.}~\bibnamefont {Yunoki}},\
  and\ \bibinfo {author} {\bibfnamefont {S.}~\bibnamefont {Sorella}},\
  }\bibfield  {title} {\bibinfo {title} {Universal {{Quantum Criticality}} in
  the {{Metal-Insulator Transition}} of {{Two-Dimensional Interacting Dirac
  Electrons}}},\ }\href {https://doi.org/10.1103/PhysRevX.6.011029} {\bibfield
  {journal} {\bibinfo  {journal} {Phys. Rev. X}\ }\textbf {\bibinfo {volume}
  {6}},\ \bibinfo {pages} {011029} (\bibinfo {year} {2016})}\BibitemShut
  {NoStop}%
\bibitem [{\citenamefont {Schwierz}(2010)}]{Schwierz-NatNanotech2010}%
  \BibitemOpen
  \bibfield  {author} {\bibinfo {author} {\bibfnamefont {F.}~\bibnamefont
  {Schwierz}},\ }\bibfield  {title} {\bibinfo {title} {Graphene transistors},\
  }\href {https://doi.org/10.1038/nnano.2010.89} {\bibfield  {journal}
  {\bibinfo  {journal} {Nat. Nanotechnol.}\ }\textbf {\bibinfo {volume} {5}},\
  \bibinfo {pages} {487} (\bibinfo {year} {2010})}\BibitemShut {NoStop}%
\bibitem [{\citenamefont {Reich}\ \emph {et~al.}(2002)\citenamefont {Reich},
  \citenamefont {Maultzsch}, \citenamefont {Thomsen},\ and\ \citenamefont
  {Ordej{\'o}n}}]{Reich-PRB2002}%
  \BibitemOpen
  \bibfield  {author} {\bibinfo {author} {\bibfnamefont {S.}~\bibnamefont
  {Reich}}, \bibinfo {author} {\bibfnamefont {J.}~\bibnamefont {Maultzsch}},
  \bibinfo {author} {\bibfnamefont {C.}~\bibnamefont {Thomsen}},\ and\ \bibinfo
  {author} {\bibfnamefont {P.}~\bibnamefont {Ordej{\'o}n}},\ }\bibfield
  {title} {\bibinfo {title} {Tight-binding description of graphene},\ }\href
  {https://doi.org/10.1103/PhysRevB.66.035412} {\bibfield  {journal} {\bibinfo
  {journal} {Phys. Rev. B}\ }\textbf {\bibinfo {volume} {66}},\ \bibinfo
  {pages} {035412} (\bibinfo {year} {2002})}\BibitemShut {NoStop}%
\bibitem [{\citenamefont {Jung}\ and\ \citenamefont
  {MacDonald}(2013)}]{Jung-PRB2013}%
  \BibitemOpen
  \bibfield  {author} {\bibinfo {author} {\bibfnamefont {J.}~\bibnamefont
  {Jung}}\ and\ \bibinfo {author} {\bibfnamefont {A.~H.}\ \bibnamefont
  {MacDonald}},\ }\bibfield  {title} {\bibinfo {title} {Tight-binding model for
  graphene {$\pi$}-bands from maximally localized wannier functions},\ }\href
  {https://doi.org/10.1103/PhysRevB.87.195450} {\bibfield  {journal} {\bibinfo
  {journal} {Phys. Rev. B}\ }\textbf {\bibinfo {volume} {87}},\ \bibinfo
  {pages} {195450} (\bibinfo {year} {2013})}\BibitemShut {NoStop}%
\bibitem [{\citenamefont {Yazyev}(2008)}]{Yazyev-PRL2008}%
  \BibitemOpen
  \bibfield  {author} {\bibinfo {author} {\bibfnamefont {O.~V.}\ \bibnamefont
  {Yazyev}},\ }\bibfield  {title} {\bibinfo {title} {Magnetism in {{Disordered
  Graphene}} and {{Irradiated Graphite}}},\ }\href
  {https://doi.org/10.1103/PhysRevLett.101.037203} {\bibfield  {journal}
  {\bibinfo  {journal} {Phys. Rev. Lett.}\ }\textbf {\bibinfo {volume} {101}},\
  \bibinfo {pages} {037203} (\bibinfo {year} {2008})}\BibitemShut {NoStop}%
\bibitem [{\citenamefont {Dutta}\ \emph {et~al.}(2008)\citenamefont {Dutta},
  \citenamefont {Lakshmi},\ and\ \citenamefont {Pati}}]{Dutta-PRB2008}%
  \BibitemOpen
  \bibfield  {author} {\bibinfo {author} {\bibfnamefont {S.}~\bibnamefont
  {Dutta}}, \bibinfo {author} {\bibfnamefont {S.}~\bibnamefont {Lakshmi}},\
  and\ \bibinfo {author} {\bibfnamefont {S.~K.}\ \bibnamefont {Pati}},\
  }\bibfield  {title} {\bibinfo {title} {Electron-electron interactions on the
  edge states of graphene: {{A}} many-body configuration interaction study},\
  }\href {https://doi.org/10.1103/PhysRevB.77.073412} {\bibfield  {journal}
  {\bibinfo  {journal} {Phys. Rev. B}\ }\textbf {\bibinfo {volume} {77}},\
  \bibinfo {pages} {073412} (\bibinfo {year} {2008})}\BibitemShut {NoStop}%
\bibitem [{\citenamefont {Wehling}\ \emph {et~al.}(2011)\citenamefont
  {Wehling}, \citenamefont {{{\c{S}}a{\c{s}}{{\i}}o{\u{g}}lu}}, \citenamefont
  {Friedrich}, \citenamefont {Lichtenstein}, \citenamefont {Katsnelson},\ and\
  \citenamefont {Bl{\"{u}}gel}}]{Wehling-PRL2011}%
  \BibitemOpen
  \bibfield  {author} {\bibinfo {author} {\bibfnamefont {T.~O.}\ \bibnamefont
  {Wehling}}, \bibinfo {author} {\bibfnamefont {E.}~\bibnamefont
  {{{\c{S}}a{\c{s}}{{\i}}o{\u{g}}lu}}}, \bibinfo {author} {\bibfnamefont
  {C.}~\bibnamefont {Friedrich}}, \bibinfo {author} {\bibfnamefont {A.~I.}\
  \bibnamefont {Lichtenstein}}, \bibinfo {author} {\bibfnamefont {M.~I.}\
  \bibnamefont {Katsnelson}},\ and\ \bibinfo {author} {\bibfnamefont
  {S.}~\bibnamefont {Bl{\"{u}}gel}},\ }\bibfield  {title} {\bibinfo {title}
  {Strength of effective coulomb interactions in graphene and graphite},\
  }\href {https://doi.org/10.1103/PhysRevLett.106.236805} {\bibfield  {journal}
  {\bibinfo  {journal} {Phys. Rev. Lett.}\ }\textbf {\bibinfo {volume} {106}},\
  \bibinfo {pages} {236805} (\bibinfo {year} {2011})}\BibitemShut {NoStop}%
\bibitem [{\citenamefont {Jung}\ and\ \citenamefont
  {MacDonald}(2011)}]{Jung-PRB2011}%
  \BibitemOpen
  \bibfield  {author} {\bibinfo {author} {\bibfnamefont {J.}~\bibnamefont
  {Jung}}\ and\ \bibinfo {author} {\bibfnamefont {A.~H.}\ \bibnamefont
  {MacDonald}},\ }\bibfield  {title} {\bibinfo {title} {Enhancement of nonlocal
  exchange near isolated band crossings in graphene},\ }\href
  {https://doi.org/10.1103/PhysRevB.84.085446} {\bibfield  {journal} {\bibinfo
  {journal} {Phys. Rev. B}\ }\textbf {\bibinfo {volume} {84}},\ \bibinfo
  {pages} {085446} (\bibinfo {year} {2011})}\BibitemShut {NoStop}%
\bibitem [{\citenamefont {Tang}\ \emph {et~al.}(2015)\citenamefont {Tang},
  \citenamefont {Laksono}, \citenamefont {Rodrigues}, \citenamefont {Sengupta},
  \citenamefont {Assaad},\ and\ \citenamefont {Adam}}]{Tang-PRL2015}%
  \BibitemOpen
  \bibfield  {author} {\bibinfo {author} {\bibfnamefont {H.~K.}\ \bibnamefont
  {Tang}}, \bibinfo {author} {\bibfnamefont {E.}~\bibnamefont {Laksono}},
  \bibinfo {author} {\bibfnamefont {J.~N.~B.}\ \bibnamefont {Rodrigues}},
  \bibinfo {author} {\bibfnamefont {P.}~\bibnamefont {Sengupta}}, \bibinfo
  {author} {\bibfnamefont {F.~F.}\ \bibnamefont {Assaad}},\ and\ \bibinfo
  {author} {\bibfnamefont {S.}~\bibnamefont {Adam}},\ }\bibfield  {title}
  {\bibinfo {title} {Interaction-{{Driven Metal-Insulator Transition}} in
  {{Strained Graphene}}},\ }\href
  {https://doi.org/10.1103/PhysRevLett.115.186602} {\bibfield  {journal}
  {\bibinfo  {journal} {Phys. Rev. Lett.}\ }\textbf {\bibinfo {volume} {115}},\
  \bibinfo {pages} {186602} (\bibinfo {year} {2015})}\BibitemShut {NoStop}%
\bibitem [{\citenamefont {Seki}\ \emph {et~al.}(2019)\citenamefont {Seki},
  \citenamefont {Otsuka}, \citenamefont {Yunoki},\ and\ \citenamefont
  {Sorella}}]{Seki-PRB2019}%
  \BibitemOpen
  \bibfield  {author} {\bibinfo {author} {\bibfnamefont {K.}~\bibnamefont
  {Seki}}, \bibinfo {author} {\bibfnamefont {Y.}~\bibnamefont {Otsuka}},
  \bibinfo {author} {\bibfnamefont {S.}~\bibnamefont {Yunoki}},\ and\ \bibinfo
  {author} {\bibfnamefont {S.}~\bibnamefont {Sorella}},\ }\bibfield  {title}
  {\bibinfo {title} {Fermi-liquid ground state of interacting {{Dirac}}
  fermions in two dimensions},\ }\href
  {https://doi.org/10.1103/PhysRevB.99.125145} {\bibfield  {journal} {\bibinfo
  {journal} {Phys. Rev. B}\ }\textbf {\bibinfo {volume} {99}},\ \bibinfo
  {pages} {125145} (\bibinfo {year} {2019})}\BibitemShut {NoStop}%
\bibitem [{\citenamefont {Raczkowski}\ \emph {et~al.}(2020)\citenamefont
  {Raczkowski}, \citenamefont {Peters}, \citenamefont {Ph{\`u}ng},
  \citenamefont {Takemori}, \citenamefont {Assaad}, \citenamefont {Honecker},\
  and\ \citenamefont {Vahedi}}]{Raczkowski-PRB2020}%
  \BibitemOpen
  \bibfield  {author} {\bibinfo {author} {\bibfnamefont {M.}~\bibnamefont
  {Raczkowski}}, \bibinfo {author} {\bibfnamefont {R.}~\bibnamefont {Peters}},
  \bibinfo {author} {\bibfnamefont {T.~T.}\ \bibnamefont {Ph{\`u}ng}}, \bibinfo
  {author} {\bibfnamefont {N.}~\bibnamefont {Takemori}}, \bibinfo {author}
  {\bibfnamefont {F.~F.}\ \bibnamefont {Assaad}}, \bibinfo {author}
  {\bibfnamefont {A.}~\bibnamefont {Honecker}},\ and\ \bibinfo {author}
  {\bibfnamefont {J.}~\bibnamefont {Vahedi}},\ }\bibfield  {title} {\bibinfo
  {title} {Hubbard model on the honeycomb lattice: {{From}} static and
  dynamical mean-field theories to lattice quantum {{Monte Carlo}}
  simulations},\ }\href {https://doi.org/10.1103/PhysRevB.101.125103}
  {\bibfield  {journal} {\bibinfo  {journal} {Phys. Rev. B}\ }\textbf {\bibinfo
  {volume} {101}},\ \bibinfo {pages} {125103} (\bibinfo {year}
  {2020})}\BibitemShut {NoStop}%
\bibitem [{\citenamefont {Ostmeyer}\ \emph {et~al.}(2021)\citenamefont
  {Ostmeyer}, \citenamefont {Berkowitz}, \citenamefont {Krieg}, \citenamefont
  {L{\"a}hde}, \citenamefont {Luu},\ and\ \citenamefont
  {Urbach}}]{Ostmeyer-PRB2021}%
  \BibitemOpen
  \bibfield  {author} {\bibinfo {author} {\bibfnamefont {J.}~\bibnamefont
  {Ostmeyer}}, \bibinfo {author} {\bibfnamefont {E.}~\bibnamefont {Berkowitz}},
  \bibinfo {author} {\bibfnamefont {S.}~\bibnamefont {Krieg}}, \bibinfo
  {author} {\bibfnamefont {T.~A.}\ \bibnamefont {L{\"a}hde}}, \bibinfo {author}
  {\bibfnamefont {T.}~\bibnamefont {Luu}},\ and\ \bibinfo {author}
  {\bibfnamefont {C.}~\bibnamefont {Urbach}},\ }\bibfield  {title} {\bibinfo
  {title} {Antiferromagnetic character of the quantum phase transition in the
  {{Hubbard}} model on the honeycomb lattice},\ }\href
  {https://doi.org/10.1103/PhysRevB.104.155142} {\bibfield  {journal} {\bibinfo
   {journal} {Phys. Rev. B}\ }\textbf {\bibinfo {volume} {104}},\ \bibinfo
  {pages} {155142} (\bibinfo {year} {2021})}\BibitemShut {NoStop}%
\bibitem [{\citenamefont {Guinea}\ \emph {et~al.}(2010)\citenamefont {Guinea},
  \citenamefont {Katsnelson},\ and\ \citenamefont {Geim}}]{Guinea-NatPhys2010}%
  \BibitemOpen
  \bibfield  {author} {\bibinfo {author} {\bibfnamefont {F.}~\bibnamefont
  {Guinea}}, \bibinfo {author} {\bibfnamefont {M.~I.}\ \bibnamefont
  {Katsnelson}},\ and\ \bibinfo {author} {\bibfnamefont {A.~K.}\ \bibnamefont
  {Geim}},\ }\bibfield  {title} {\bibinfo {title} {Energy gaps and a zero-field
  quantum {{Hall}} effect in graphene by strain engineering},\ }\href
  {https://doi.org/10.1038/nphys1420} {\bibfield  {journal} {\bibinfo
  {journal} {Nat. Phys.}\ }\textbf {\bibinfo {volume} {6}},\ \bibinfo {pages}
  {30} (\bibinfo {year} {2010})}\BibitemShut {NoStop}%
\bibitem [{\citenamefont {Choi}\ \emph {et~al.}(2010)\citenamefont {Choi},
  \citenamefont {Jhi},\ and\ \citenamefont {Son}}]{Choi-PRB2010}%
  \BibitemOpen
  \bibfield  {author} {\bibinfo {author} {\bibfnamefont {S.-M.}\ \bibnamefont
  {Choi}}, \bibinfo {author} {\bibfnamefont {S.-H.}\ \bibnamefont {Jhi}},\ and\
  \bibinfo {author} {\bibfnamefont {Y.-W.}\ \bibnamefont {Son}},\ }\bibfield
  {title} {\bibinfo {title} {Effects of strain on electronic properties of
  graphene},\ }\href {https://doi.org/10.1103/PhysRevB.81.081407} {\bibfield
  {journal} {\bibinfo  {journal} {Phys. Rev. B}\ }\textbf {\bibinfo {volume}
  {81}},\ \bibinfo {pages} {081407(R)} (\bibinfo {year} {2010})}\BibitemShut
  {NoStop}%
\bibitem [{\citenamefont {Cocco}\ \emph {et~al.}(2010)\citenamefont {Cocco},
  \citenamefont {Cadelano},\ and\ \citenamefont {Colombo}}]{Cocco-PRB2010}%
  \BibitemOpen
  \bibfield  {author} {\bibinfo {author} {\bibfnamefont {G.}~\bibnamefont
  {Cocco}}, \bibinfo {author} {\bibfnamefont {E.}~\bibnamefont {Cadelano}},\
  and\ \bibinfo {author} {\bibfnamefont {L.}~\bibnamefont {Colombo}},\
  }\bibfield  {title} {\bibinfo {title} {Gap opening in graphene by shear
  strain},\ }\href {https://doi.org/10.1103/PhysRevB.81.241412} {\bibfield
  {journal} {\bibinfo  {journal} {Phys. Rev. B}\ }\textbf {\bibinfo {volume}
  {81}},\ \bibinfo {pages} {241412(R)} (\bibinfo {year} {2010})}\BibitemShut
  {NoStop}%
\bibitem [{\citenamefont {Jiang}\ \emph {et~al.}(2017)\citenamefont {Jiang},
  \citenamefont {Mao}, \citenamefont {Duan}, \citenamefont {Lai}, \citenamefont
  {Watanabe}, \citenamefont {Taniguchi},\ and\ \citenamefont
  {Andrei}}]{Jiang-NanoLett2017}%
  \BibitemOpen
  \bibfield  {author} {\bibinfo {author} {\bibfnamefont {Y.}~\bibnamefont
  {Jiang}}, \bibinfo {author} {\bibfnamefont {J.}~\bibnamefont {Mao}}, \bibinfo
  {author} {\bibfnamefont {J.}~\bibnamefont {Duan}}, \bibinfo {author}
  {\bibfnamefont {X.}~\bibnamefont {Lai}}, \bibinfo {author} {\bibfnamefont
  {K.}~\bibnamefont {Watanabe}}, \bibinfo {author} {\bibfnamefont
  {T.}~\bibnamefont {Taniguchi}},\ and\ \bibinfo {author} {\bibfnamefont
  {E.~Y.}\ \bibnamefont {Andrei}},\ }\bibfield  {title} {\bibinfo {title}
  {Visualizing {{Strain-Induced Pseudomagnetic Fields}} in {{Graphene}} through
  an {{hBN Magnifying Glass}}},\ }\href
  {https://doi.org/10.1021/acs.nanolett.6b05228} {\bibfield  {journal}
  {\bibinfo  {journal} {Nano Letters}\ }\textbf {\bibinfo {volume} {17}},\
  \bibinfo {pages} {2839} (\bibinfo {year} {2017})}\BibitemShut {NoStop}%
\bibitem [{\citenamefont {Si}\ \emph {et~al.}(2016)\citenamefont {Si},
  \citenamefont {Sun},\ and\ \citenamefont {Liu}}]{Si-Nanoiscale2016}%
  \BibitemOpen
  \bibfield  {author} {\bibinfo {author} {\bibfnamefont {C.}~\bibnamefont
  {Si}}, \bibinfo {author} {\bibfnamefont {Z.}~\bibnamefont {Sun}},\ and\
  \bibinfo {author} {\bibfnamefont {F.}~\bibnamefont {Liu}},\ }\bibfield
  {title} {\bibinfo {title} {Strain engineering of graphene: {{A}} review},\
  }\href {https://doi.org/10.1039/c5nr07755a} {\bibfield  {journal} {\bibinfo
  {journal} {Nanoscale}\ }\textbf {\bibinfo {volume} {8}},\ \bibinfo {pages}
  {3207} (\bibinfo {year} {2016})}\BibitemShut {NoStop}%
\bibitem [{\citenamefont {Naumis}\ \emph {et~al.}(2017)\citenamefont {Naumis},
  \citenamefont {{Barraza-Lopez}}, \citenamefont {{Oliva-Leyva}},\ and\
  \citenamefont {Terrones}}]{Naumis-RepProgPhys2017}%
  \BibitemOpen
  \bibfield  {author} {\bibinfo {author} {\bibfnamefont {G.~G.}\ \bibnamefont
  {Naumis}}, \bibinfo {author} {\bibfnamefont {S.}~\bibnamefont
  {{Barraza-Lopez}}}, \bibinfo {author} {\bibfnamefont {M.}~\bibnamefont
  {{Oliva-Leyva}}},\ and\ \bibinfo {author} {\bibfnamefont {H.}~\bibnamefont
  {Terrones}},\ }\bibfield  {title} {\bibinfo {title} {Electronic and optical
  properties of strained graphene and other strained {{2D}} materials: {{A}}
  review},\ }\href {https://doi.org/10.1088/1361-6633/aa74ef} {\bibfield
  {journal} {\bibinfo  {journal} {Rep. Prog. Phys.}\ }\textbf {\bibinfo
  {volume} {80}},\ \bibinfo {pages} {096501} (\bibinfo {year}
  {2017})}\BibitemShut {NoStop}%
\bibitem [{\citenamefont {Sorella}\ \emph {et~al.}(2018)\citenamefont
  {Sorella}, \citenamefont {Seki}, \citenamefont {Brovko}, \citenamefont
  {Shirakawa}, \citenamefont {Miyakoshi}, \citenamefont {Yunoki},\ and\
  \citenamefont {Tosatti}}]{Sorella-PRL2018}%
  \BibitemOpen
  \bibfield  {author} {\bibinfo {author} {\bibfnamefont {S.}~\bibnamefont
  {Sorella}}, \bibinfo {author} {\bibfnamefont {K.}~\bibnamefont {Seki}},
  \bibinfo {author} {\bibfnamefont {O.~O.}\ \bibnamefont {Brovko}}, \bibinfo
  {author} {\bibfnamefont {T.}~\bibnamefont {Shirakawa}}, \bibinfo {author}
  {\bibfnamefont {S.}~\bibnamefont {Miyakoshi}}, \bibinfo {author}
  {\bibfnamefont {S.}~\bibnamefont {Yunoki}},\ and\ \bibinfo {author}
  {\bibfnamefont {E.}~\bibnamefont {Tosatti}},\ }\bibfield  {title} {\bibinfo
  {title} {Correlation-{{Driven Dimerization}} and {{Topological Gap Opening}}
  in {{Isotropically Strained Graphene}}},\ }\href
  {https://doi.org/10.1103/PhysRevLett.121.066402} {\bibfield  {journal}
  {\bibinfo  {journal} {Phys. Rev. Lett.}\ }\textbf {\bibinfo {volume} {121}},\
  \bibinfo {pages} {066402} (\bibinfo {year} {2018})}\BibitemShut {NoStop}%
\bibitem [{\citenamefont {Lee}\ \emph {et~al.}(2012)\citenamefont {Lee},
  \citenamefont {Kim},\ and\ \citenamefont {Kim}}]{Lee-PRB2012}%
  \BibitemOpen
  \bibfield  {author} {\bibinfo {author} {\bibfnamefont {S.-H.}\ \bibnamefont
  {Lee}}, \bibinfo {author} {\bibfnamefont {S.}~\bibnamefont {Kim}},\ and\
  \bibinfo {author} {\bibfnamefont {K.}~\bibnamefont {Kim}},\ }\bibfield
  {title} {\bibinfo {title} {Semimetal-antiferromagnetic insulator transition
  in graphene induced by biaxial strain},\ }\href
  {https://doi.org/10.1103/PhysRevB.86.155436} {\bibfield  {journal} {\bibinfo
  {journal} {Phys. Rev. B}\ }\textbf {\bibinfo {volume} {86}},\ \bibinfo
  {pages} {155436} (\bibinfo {year} {2012})}\BibitemShut {NoStop}%
\bibitem [{\citenamefont {Tang}\ and\ \citenamefont
  {Hirsch}(1988)}]{Tang-PRB1988}%
  \BibitemOpen
  \bibfield  {author} {\bibinfo {author} {\bibfnamefont {S.}~\bibnamefont
  {Tang}}\ and\ \bibinfo {author} {\bibfnamefont {J.~E.}\ \bibnamefont
  {Hirsch}},\ }\bibfield  {title} {\bibinfo {title} {Peierls instability in the
  two-dimensional half-filled {{Hubbard}} model},\ }\href
  {https://doi.org/10.1103/PhysRevB.37.9546} {\bibfield  {journal} {\bibinfo
  {journal} {Phys. Rev. B}\ }\textbf {\bibinfo {volume} {37}},\ \bibinfo
  {pages} {9546} (\bibinfo {year} {1988})}\BibitemShut {NoStop}%
\bibitem [{\citenamefont {Mazumdar}(1987)}]{Mazumdar-PRB1987}%
  \BibitemOpen
  \bibfield  {author} {\bibinfo {author} {\bibfnamefont {S.}~\bibnamefont
  {Mazumdar}},\ }\bibfield  {title} {\bibinfo {title} {Valence-bond approach to
  two-dimensional broken symmetries: Application to {{La$_{2}$CuO$_{4}$}}},\
  }\href {https://doi.org/10.1103/PhysRevB.36.7190} {\bibfield  {journal}
  {\bibinfo  {journal} {Phys. Rev. B}\ }\textbf {\bibinfo {volume} {36}},\
  \bibinfo {pages} {7190} (\bibinfo {year} {1987})}\BibitemShut {NoStop}%
\bibitem [{\citenamefont {Ono}\ and\ \citenamefont
  {Hamano}(2000)}]{Ono-JPSJ2000}%
  \BibitemOpen
  \bibfield  {author} {\bibinfo {author} {\bibfnamefont {Y.}~\bibnamefont
  {Ono}}\ and\ \bibinfo {author} {\bibfnamefont {T.}~\bibnamefont {Hamano}},\
  }\bibfield  {title} {\bibinfo {title} {Peierls {{Distortion}} in
  {{Two-Dimensional Tight-Binding Model}}},\ }\href
  {https://doi.org/10.1143/JPSJ.69.1769} {\bibfield  {journal} {\bibinfo
  {journal} {J. Phys. Soc. Jpn.}\ }\textbf {\bibinfo {volume} {69}},\ \bibinfo
  {pages} {1769} (\bibinfo {year} {2000})}\BibitemShut {NoStop}%
\bibitem [{\citenamefont {Chiba}\ and\ \citenamefont
  {Ono}(2003)}]{Chiba-JPSJ2003}%
  \BibitemOpen
  \bibfield  {author} {\bibinfo {author} {\bibfnamefont {S.}~\bibnamefont
  {Chiba}}\ and\ \bibinfo {author} {\bibfnamefont {Y.}~\bibnamefont {Ono}},\
  }\bibfield  {title} {\bibinfo {title} {Multi {{Mode Phonon Softening}} in
  {{Two-Dimensional Electron}}{\textendash}{{Lattice System}}},\ }\href
  {https://doi.org/10.1143/JPSJ.72.1995} {\bibfield  {journal} {\bibinfo
  {journal} {J. Phys. Soc. Jpn.}\ }\textbf {\bibinfo {volume} {72}},\ \bibinfo
  {pages} {1995} (\bibinfo {year} {2003})}\BibitemShut {NoStop}%
\bibitem [{\citenamefont {Chiba}\ and\ \citenamefont
  {Ono}(2004)}]{Chiba-JPSJ2004a}%
  \BibitemOpen
  \bibfield  {author} {\bibinfo {author} {\bibfnamefont {S.}~\bibnamefont
  {Chiba}}\ and\ \bibinfo {author} {\bibfnamefont {Y.}~\bibnamefont {Ono}},\
  }\bibfield  {title} {\bibinfo {title} {Phonon {{Dispersion Relations}} in
  {{Two-dimensional Peierls Phase}}},\ }\href
  {https://doi.org/10.1143/JPSJ.73.2473} {\bibfield  {journal} {\bibinfo
  {journal} {J. Phys. Soc. Jpn.}\ }\textbf {\bibinfo {volume} {73}},\ \bibinfo
  {pages} {2473} (\bibinfo {year} {2004})}\BibitemShut {NoStop}%
\bibitem [{\citenamefont {Xing}\ \emph {et~al.}(2021)\citenamefont {Xing},
  \citenamefont {Chiu}, \citenamefont {Poletti}, \citenamefont {Scalettar},\
  and\ \citenamefont {Batrouni}}]{Xing-PRL2021}%
  \BibitemOpen
  \bibfield  {author} {\bibinfo {author} {\bibfnamefont {B.}~\bibnamefont
  {Xing}}, \bibinfo {author} {\bibfnamefont {W.~T.}\ \bibnamefont {Chiu}},
  \bibinfo {author} {\bibfnamefont {D.}~\bibnamefont {Poletti}}, \bibinfo
  {author} {\bibfnamefont {R.~T.}\ \bibnamefont {Scalettar}},\ and\ \bibinfo
  {author} {\bibfnamefont {G.}~\bibnamefont {Batrouni}},\ }\bibfield  {title}
  {\bibinfo {title} {Quantum {{Monte Carlo Simulations}} of the {{2D
  Su-Schrieffer-Heeger Model}}},\ }\href
  {https://doi.org/10.1103/PhysRevLett.126.017601} {\bibfield  {journal}
  {\bibinfo  {journal} {Phys. Rev. Lett.}\ }\textbf {\bibinfo {volume} {126}},\
  \bibinfo {pages} {017601} (\bibinfo {year} {2021})}\BibitemShut {NoStop}%
\bibitem [{\citenamefont {Frank}\ and\ \citenamefont
  {Lieb}(2011)}]{Frank-PRL2011}%
  \BibitemOpen
  \bibfield  {author} {\bibinfo {author} {\bibfnamefont {R.~L.}\ \bibnamefont
  {Frank}}\ and\ \bibinfo {author} {\bibfnamefont {E.~H.}\ \bibnamefont
  {Lieb}},\ }\bibfield  {title} {\bibinfo {title} {Possible lattice distortions
  in the {{Hubbard}} model for graphene},\ }\href
  {https://doi.org/10.1103/PhysRevLett.107.066801} {\bibfield  {journal}
  {\bibinfo  {journal} {Phys. Rev. Lett.}\ }\textbf {\bibinfo {volume} {107}},\
  \bibinfo {pages} {066801} (\bibinfo {year} {2011})}\BibitemShut {NoStop}%
\bibitem [{\citenamefont {Blankenbecler}\ \emph {et~al.}(1981)\citenamefont
  {Blankenbecler}, \citenamefont {Scalapino},\ and\ \citenamefont
  {Sugar}}]{Blankenbecler-PRD1981}%
  \BibitemOpen
  \bibfield  {author} {\bibinfo {author} {\bibfnamefont {R.}~\bibnamefont
  {Blankenbecler}}, \bibinfo {author} {\bibfnamefont {D.~J.}\ \bibnamefont
  {Scalapino}},\ and\ \bibinfo {author} {\bibfnamefont {R.~L.}\ \bibnamefont
  {Sugar}},\ }\bibfield  {title} {\bibinfo {title} {Monte {{Carlo}}
  calculations of coupled boson-fermion systems. {{I}}},\ }\href
  {https://doi.org/10.1103/PhysRevD.24.2278} {\bibfield  {journal} {\bibinfo
  {journal} {Phys. Rev. D}\ }\textbf {\bibinfo {volume} {24}},\ \bibinfo
  {pages} {2278} (\bibinfo {year} {1981})}\BibitemShut {NoStop}%
\bibitem [{\citenamefont {Hirsch}(1985)}]{Hirsch-PRB1985}%
  \BibitemOpen
  \bibfield  {author} {\bibinfo {author} {\bibfnamefont {J.~E.}\ \bibnamefont
  {Hirsch}},\ }\bibfield  {title} {\bibinfo {title} {Two-dimensional
  {{Hubbard}} model: {{Numerical}} simulation study},\ }\href
  {https://doi.org/10.1103/PhysRevB.31.4403} {\bibfield  {journal} {\bibinfo
  {journal} {Phys. Rev. B}\ }\textbf {\bibinfo {volume} {31}},\ \bibinfo
  {pages} {4403} (\bibinfo {year} {1985})}\BibitemShut {NoStop}%
\bibitem [{\citenamefont {White}\ \emph {et~al.}(1989)\citenamefont {White},
  \citenamefont {Scalapino}, \citenamefont {Sugar}, \citenamefont {Loh},
  \citenamefont {Gubernatis},\ and\ \citenamefont {Scalettar}}]{White-PRB1989}%
  \BibitemOpen
  \bibfield  {author} {\bibinfo {author} {\bibfnamefont {S.~R.}\ \bibnamefont
  {White}}, \bibinfo {author} {\bibfnamefont {D.~J.}\ \bibnamefont
  {Scalapino}}, \bibinfo {author} {\bibfnamefont {R.~L.}\ \bibnamefont
  {Sugar}}, \bibinfo {author} {\bibfnamefont {E.~Y.}\ \bibnamefont {Loh}},
  \bibinfo {author} {\bibfnamefont {J.~E.}\ \bibnamefont {Gubernatis}},\ and\
  \bibinfo {author} {\bibfnamefont {R.~T.}\ \bibnamefont {Scalettar}},\
  }\bibfield  {title} {\bibinfo {title} {Numerical study of the two-dimensional
  {{Hubbard}} model},\ }\href {https://doi.org/10.1103/PhysRevB.40.506}
  {\bibfield  {journal} {\bibinfo  {journal} {Phys. Rev. B}\ }\textbf {\bibinfo
  {volume} {40}},\ \bibinfo {pages} {506} (\bibinfo {year} {1989})}\BibitemShut
  {NoStop}%
\bibitem [{\citenamefont {Assaad}\ and\ \citenamefont
  {Evertz}(2008)}]{Assaad-CMP2008}%
  \BibitemOpen
  \bibfield  {author} {\bibinfo {author} {\bibfnamefont {F.}~\bibnamefont
  {Assaad}}\ and\ \bibinfo {author} {\bibfnamefont {H.}~\bibnamefont
  {Evertz}},\ }\bibfield  {title} {\bibinfo {title} {World-line and
  {{Determinantal Quantum Monte Carlo Methods}} for {{Spins}}, {{Phonons}} and
  {{Electrons}}},\ }in\ \href {https://doi.org/10.1007/978-3-540-74686-7_10}
  {\emph {\bibinfo {booktitle} {Computational {{Many-Particle Physics}}}}},\
  \bibinfo {editor} {edited by\ \bibinfo {editor} {\bibfnamefont
  {H.}~\bibnamefont {Fehske}}, \bibinfo {editor} {\bibfnamefont
  {R.}~\bibnamefont {Schneider}},\ and\ \bibinfo {editor} {\bibfnamefont
  {A.}~\bibnamefont {Wei{\ss}e}}}\ (\bibinfo  {publisher} {{Springer}},\
  \bibinfo {address} {{Berlin}},\ \bibinfo {year} {2008})\ p.\ \bibinfo {pages}
  {277}\BibitemShut {NoStop}%
\bibitem [{\citenamefont {Sandvik}\ and\ \citenamefont
  {Kurkij{\"a}rvi}(1991)}]{Sandvik-PRB1991}%
  \BibitemOpen
  \bibfield  {author} {\bibinfo {author} {\bibfnamefont {A.~W.}\ \bibnamefont
  {Sandvik}}\ and\ \bibinfo {author} {\bibfnamefont {J.}~\bibnamefont
  {Kurkij{\"a}rvi}},\ }\bibfield  {title} {\bibinfo {title} {Quantum {{Monte
  Carlo}} simulation method for spin systems},\ }\href
  {https://doi.org/10.1103/PhysRevB.43.5950} {\bibfield  {journal} {\bibinfo
  {journal} {Phys. Rev. B}\ }\textbf {\bibinfo {volume} {43}},\ \bibinfo
  {pages} {5950} (\bibinfo {year} {1991})}\BibitemShut {NoStop}%
\bibitem [{\citenamefont {Sandvik}(1992)}]{Sandvik-JPA1992}%
  \BibitemOpen
  \bibfield  {author} {\bibinfo {author} {\bibfnamefont {A.~W.}\ \bibnamefont
  {Sandvik}},\ }\bibfield  {title} {\bibinfo {title} {A generalization of
  {{Handscomb}}'s quantum {{Monte Carlo}} scheme-application to the {{1D
  Hubbard}} model},\ }\href
  {https://doi.org/http://dx.doi.org/10.1088/0305-4470/25/13/017} {\bibfield
  {journal} {\bibinfo  {journal} {J. Phys. A: Math. Gen.}\ }\textbf {\bibinfo
  {volume} {25}},\ \bibinfo {pages} {3667} (\bibinfo {year}
  {1992})}\BibitemShut {NoStop}%
\bibitem [{\citenamefont {Sandvik}(1999)}]{Sandvik-PRB1999}%
  \BibitemOpen
  \bibfield  {author} {\bibinfo {author} {\bibfnamefont {A.~W.}\ \bibnamefont
  {Sandvik}},\ }\bibfield  {title} {\bibinfo {title} {Stochastic series
  expansion method with operator-loop update},\ }\href
  {https://doi.org/10.1103/PhysRevB.59.R14157} {\bibfield  {journal} {\bibinfo
  {journal} {Phys. Rev. B}\ }\textbf {\bibinfo {volume} {59}},\ \bibinfo
  {pages} {R14157} (\bibinfo {year} {1999})}\BibitemShut {NoStop}%
\bibitem [{\citenamefont {Otsuka}\ \emph {et~al.}(2008)\citenamefont {Otsuka},
  \citenamefont {Seo}, \citenamefont {Motome},\ and\ \citenamefont
  {Kato}}]{Otsuka-JPSJ2008}%
  \BibitemOpen
  \bibfield  {author} {\bibinfo {author} {\bibfnamefont {Y.}~\bibnamefont
  {Otsuka}}, \bibinfo {author} {\bibfnamefont {H.}~\bibnamefont {Seo}},
  \bibinfo {author} {\bibfnamefont {Y.}~\bibnamefont {Motome}},\ and\ \bibinfo
  {author} {\bibfnamefont {T.}~\bibnamefont {Kato}},\ }\bibfield  {title}
  {\bibinfo {title} {Finite-temperature phase diagram of quasi-one-dimensional
  molecular conductors: {{Quantum Monte Carlo}} study},\ }\href
  {https://doi.org/10.1143/JPSJ.77.113705} {\bibfield  {journal} {\bibinfo
  {journal} {J. Phys. Soc. Jpn.}\ }\textbf {\bibinfo {volume} {77}},\ \bibinfo
  {pages} {113705} (\bibinfo {year} {2008})}\BibitemShut {NoStop}%
\bibitem [{\citenamefont {Otsuka}\ \emph {et~al.}(2009)\citenamefont {Otsuka},
  \citenamefont {Seo}, \citenamefont {Motome},\ and\ \citenamefont
  {Kato}}]{Otsuka-Physica2009}%
  \BibitemOpen
  \bibfield  {author} {\bibinfo {author} {\bibfnamefont {Y.}~\bibnamefont
  {Otsuka}}, \bibinfo {author} {\bibfnamefont {H.}~\bibnamefont {Seo}},
  \bibinfo {author} {\bibfnamefont {Y.}~\bibnamefont {Motome}},\ and\ \bibinfo
  {author} {\bibfnamefont {T.}~\bibnamefont {Kato}},\ }\bibfield  {title}
  {\bibinfo {title} {Phase competitions and coexistences in
  quasi-one-dimensional molecular conductors: {{Exact}} diagonalization
  study},\ }\href {https://doi.org/10.1016/j.physb.2008.11.060} {\bibfield
  {journal} {\bibinfo  {journal} {Physica B}\ }\textbf {\bibinfo {volume}
  {404}},\ \bibinfo {pages} {479} (\bibinfo {year} {2009})}\BibitemShut
  {NoStop}%
\bibitem [{\citenamefont {Otsuka}\ \emph {et~al.}(2012)\citenamefont {Otsuka},
  \citenamefont {Seo}, \citenamefont {Yoshimi},\ and\ \citenamefont
  {Kato}}]{Otsuka-Physica2012}%
  \BibitemOpen
  \bibfield  {author} {\bibinfo {author} {\bibfnamefont {Y.}~\bibnamefont
  {Otsuka}}, \bibinfo {author} {\bibfnamefont {H.}~\bibnamefont {Seo}},
  \bibinfo {author} {\bibfnamefont {K.}~\bibnamefont {Yoshimi}},\ and\ \bibinfo
  {author} {\bibfnamefont {T.}~\bibnamefont {Kato}},\ }\bibfield  {title}
  {\bibinfo {title} {Finite temperature neutral-ionic transition and lattice
  dimerization in charge-transfer complexes: {{QMC}} study},\ }\href
  {https://doi.org/10.1016/j.physb.2012.01.031} {\bibfield  {journal} {\bibinfo
   {journal} {Physica B}\ }\textbf {\bibinfo {volume} {407}},\ \bibinfo {pages}
  {1793} (\bibinfo {year} {2012})}\BibitemShut {NoStop}%
\bibitem [{\citenamefont {Yoshioka}\ \emph {et~al.}(2012)\citenamefont
  {Yoshioka}, \citenamefont {Otsuka},\ and\ \citenamefont
  {Seo}}]{Yoshioka-Crystal2012}%
  \BibitemOpen
  \bibfield  {author} {\bibinfo {author} {\bibfnamefont {H.}~\bibnamefont
  {Yoshioka}}, \bibinfo {author} {\bibfnamefont {Y.}~\bibnamefont {Otsuka}},\
  and\ \bibinfo {author} {\bibfnamefont {H.}~\bibnamefont {Seo}},\ }\bibfield
  {title} {\bibinfo {title} {Theoretical {{Studies}} on {{Phase Transitions}}
  in {{Quasi-One-Dimensional Molecular Conductors}}},\ }\href
  {https://doi.org/10.3390/cryst2030996} {\bibfield  {journal} {\bibinfo
  {journal} {Crystals}\ }\textbf {\bibinfo {volume} {2}},\ \bibinfo {pages}
  {996} (\bibinfo {year} {2012})}\BibitemShut {NoStop}%
\bibitem [{\citenamefont {Robbins}\ and\ \citenamefont
  {Monro}(1951)}]{Robbins-AMS1951}%
  \BibitemOpen
  \bibfield  {author} {\bibinfo {author} {\bibfnamefont {H.}~\bibnamefont
  {Robbins}}\ and\ \bibinfo {author} {\bibfnamefont {S.}~\bibnamefont
  {Monro}},\ }\bibfield  {title} {\bibinfo {title} {A {{Stochastic
  Approximation Method}}},\ }\href {https://doi.org/10.1214/aoms/1177729586}
  {\bibfield  {journal} {\bibinfo  {journal} {Ann. Math. Stat.}\ }\textbf
  {\bibinfo {volume} {22}},\ \bibinfo {pages} {400} (\bibinfo {year}
  {1951})}\BibitemShut {NoStop}%
\bibitem [{\citenamefont {Yasuda}\ and\ \citenamefont
  {Todo}(2013)}]{Yasuda-PRE2013}%
  \BibitemOpen
  \bibfield  {author} {\bibinfo {author} {\bibfnamefont {S.}~\bibnamefont
  {Yasuda}}\ and\ \bibinfo {author} {\bibfnamefont {S.}~\bibnamefont {Todo}},\
  }\bibfield  {title} {\bibinfo {title} {Monte {{Carlo}} simulation with
  aspect-ratio optimization: {{Anomalous}} anisotropic scaling in dimerized
  antiferromagnets},\ }\href {https://doi.org/10.1103/PhysRevE.88.061301}
  {\bibfield  {journal} {\bibinfo  {journal} {Phys. Rev. E}\ }\textbf {\bibinfo
  {volume} {88}},\ \bibinfo {pages} {061301(R)} (\bibinfo {year}
  {2013})}\BibitemShut {NoStop}%
\bibitem [{\citenamefont {Yasuda}\ \emph {et~al.}(2015)\citenamefont {Yasuda},
  \citenamefont {Suwa},\ and\ \citenamefont {Todo}}]{Yasuda-PRB2015}%
  \BibitemOpen
  \bibfield  {author} {\bibinfo {author} {\bibfnamefont {S.}~\bibnamefont
  {Yasuda}}, \bibinfo {author} {\bibfnamefont {H.}~\bibnamefont {Suwa}},\ and\
  \bibinfo {author} {\bibfnamefont {S.}~\bibnamefont {Todo}},\ }\bibfield
  {title} {\bibinfo {title} {Stochastic approximation of dynamical exponent at
  quantum critical point},\ }\href {https://doi.org/10.1103/PhysRevB.92.104411}
  {\bibfield  {journal} {\bibinfo  {journal} {Phys. Rev. B}\ }\textbf {\bibinfo
  {volume} {92}},\ \bibinfo {pages} {104411} (\bibinfo {year}
  {2015})}\BibitemShut {NoStop}%
\bibitem [{\citenamefont {Roy}\ \emph {et~al.}(2013)\citenamefont {Roy},
  \citenamefont {Juri{\v c}i{\'c}},\ and\ \citenamefont
  {Herbut}}]{Roy-PRB2013}%
  \BibitemOpen
  \bibfield  {author} {\bibinfo {author} {\bibfnamefont {B.}~\bibnamefont
  {Roy}}, \bibinfo {author} {\bibfnamefont {V.}~\bibnamefont {Juri{\v
  c}i{\'c}}},\ and\ \bibinfo {author} {\bibfnamefont {I.~F.}\ \bibnamefont
  {Herbut}},\ }\bibfield  {title} {\bibinfo {title} {Quantum superconducting
  criticality in graphene and topological insulators},\ }\href
  {https://doi.org/10.1103/PhysRevB.87.041401} {\bibfield  {journal} {\bibinfo
  {journal} {Phys. Rev. B}\ }\textbf {\bibinfo {volume} {87}},\ \bibinfo
  {pages} {041401(R)} (\bibinfo {year} {2013})}\BibitemShut {NoStop}%
\bibitem [{\citenamefont {Roy}\ and\ \citenamefont {Juri{\v
  c}i{\'c}}(2019)}]{Roy-PRB2019}%
  \BibitemOpen
  \bibfield  {author} {\bibinfo {author} {\bibfnamefont {B.}~\bibnamefont
  {Roy}}\ and\ \bibinfo {author} {\bibfnamefont {V.}~\bibnamefont {Juri{\v
  c}i{\'c}}},\ }\bibfield  {title} {\bibinfo {title} {Fermionic
  multicriticality near {{Kekul}}{\textbackslash}'e valence-bond ordering on a
  honeycomb lattice},\ }\href {https://doi.org/10.1103/PhysRevB.99.241103}
  {\bibfield  {journal} {\bibinfo  {journal} {Phys. Rev. B}\ }\textbf {\bibinfo
  {volume} {99}},\ \bibinfo {pages} {241103(R)} (\bibinfo {year}
  {2019})}\BibitemShut {NoStop}%
\bibitem [{\citenamefont {Cai}\ \emph {et~al.}(2021)\citenamefont {Cai},
  \citenamefont {Li},\ and\ \citenamefont {Yao}}]{Cai-PRL2021}%
  \BibitemOpen
  \bibfield  {author} {\bibinfo {author} {\bibfnamefont {X.}~\bibnamefont
  {Cai}}, \bibinfo {author} {\bibfnamefont {Z.~X.}\ \bibnamefont {Li}},\ and\
  \bibinfo {author} {\bibfnamefont {H.}~\bibnamefont {Yao}},\ }\bibfield
  {title} {\bibinfo {title} {Antiferromagnetism {{Induced}} by {{Bond
  Su-Schrieffer-Heeger Electron-Phonon Coupling}}: {{A Quantum Monte Carlo
  Study}}},\ }\href {https://doi.org/10.1103/PhysRevLett.127.247203} {\bibfield
   {journal} {\bibinfo  {journal} {Phys. Rev. Lett.}\ }\textbf {\bibinfo
  {volume} {127}},\ \bibinfo {pages} {247203} (\bibinfo {year}
  {2021})}\BibitemShut {NoStop}%
\bibitem [{\citenamefont {Cai}\ \emph {et~al.}(2022)\citenamefont {Cai},
  \citenamefont {Li},\ and\ \citenamefont {Yao}}]{Cai-PRB2022}%
  \BibitemOpen
  \bibfield  {author} {\bibinfo {author} {\bibfnamefont {X.}~\bibnamefont
  {Cai}}, \bibinfo {author} {\bibfnamefont {Z.-X.}\ \bibnamefont {Li}},\ and\
  \bibinfo {author} {\bibfnamefont {H.}~\bibnamefont {Yao}},\ }\bibfield
  {title} {\bibinfo {title} {Robustness of antiferromagnetism in the
  {{Su-Schrieffer-Heeger Hubbard}} model},\ }\href
  {https://doi.org/10.1103/PhysRevB.106.L081115} {\bibfield  {journal}
  {\bibinfo  {journal} {Phys. Rev. B}\ }\textbf {\bibinfo {volume} {106}},\
  \bibinfo {pages} {L081115} (\bibinfo {year} {2022})}\BibitemShut {NoStop}%
\bibitem [{\citenamefont {Feng}\ \emph {et~al.}(2022)\citenamefont {Feng},
  \citenamefont {Xing}, \citenamefont {Poletti}, \citenamefont {Scalettar},\
  and\ \citenamefont {Batrouni}}]{Feng-PRB2022}%
  \BibitemOpen
  \bibfield  {author} {\bibinfo {author} {\bibfnamefont {C.}~\bibnamefont
  {Feng}}, \bibinfo {author} {\bibfnamefont {B.}~\bibnamefont {Xing}}, \bibinfo
  {author} {\bibfnamefont {D.}~\bibnamefont {Poletti}}, \bibinfo {author}
  {\bibfnamefont {R.}~\bibnamefont {Scalettar}},\ and\ \bibinfo {author}
  {\bibfnamefont {G.}~\bibnamefont {Batrouni}},\ }\bibfield  {title} {\bibinfo
  {title} {Phase diagram of the {{Su-Schrieffer-Heeger-Hubbard}} model on a
  square lattice},\ }\href {https://doi.org/10.1103/PhysRevB.106.L081114}
  {\bibfield  {journal} {\bibinfo  {journal} {Phys. Rev. B}\ }\textbf {\bibinfo
  {volume} {106}},\ \bibinfo {pages} {L081114} (\bibinfo {year}
  {2022})}\BibitemShut {NoStop}%
\bibitem [{\citenamefont {Weber}(2021)}]{Weber-PRB2021}%
  \BibitemOpen
  \bibfield  {author} {\bibinfo {author} {\bibfnamefont {M.}~\bibnamefont
  {Weber}},\ }\bibfield  {title} {\bibinfo {title} {Valence bond order in a
  honeycomb antiferromagnet coupled to quantum phonons},\ }\href
  {https://doi.org/10.1103/PhysRevB.103.L041105} {\bibfield  {journal}
  {\bibinfo  {journal} {Phys. Rev. B}\ }\textbf {\bibinfo {volume} {103}},\
  \bibinfo {pages} {L041105} (\bibinfo {year} {2021})}\BibitemShut {NoStop}%
\bibitem [{\citenamefont {Sato}\ \emph {et~al.}(2017)\citenamefont {Sato},
  \citenamefont {Hohenadler},\ and\ \citenamefont {Assaad}}]{Sato-PRL2017}%
  \BibitemOpen
  \bibfield  {author} {\bibinfo {author} {\bibfnamefont {T.}~\bibnamefont
  {Sato}}, \bibinfo {author} {\bibfnamefont {M.}~\bibnamefont {Hohenadler}},\
  and\ \bibinfo {author} {\bibfnamefont {F.~F.}\ \bibnamefont {Assaad}},\
  }\bibfield  {title} {\bibinfo {title} {Dirac {{Fermions}} with {{Competing
  Orders}}: {{Non-Landau Transition}} with {{Emergent Symmetry}}},\ }\href
  {https://doi.org/10.1103/PhysRevLett.119.197203} {\bibfield  {journal}
  {\bibinfo  {journal} {Phys. Rev. Lett.}\ }\textbf {\bibinfo {volume} {119}},\
  \bibinfo {pages} {197203} (\bibinfo {year} {2017})}\BibitemShut {NoStop}%
\end{thebibliography}%

\end{document}